\documentclass[preprintnumbers,prd,showpacs,floatfix,superscriptaddress,nofootinbib,twocolumn, letterpaper]{revtex4-1}
\pdfoutput=1

\usepackage{longtable}
\usepackage{bm}
\usepackage{relsize}
\usepackage{amsfonts}
\usepackage{amsmath}
\usepackage{amssymb,epsf}
\usepackage{latexsym}
\usepackage{graphicx,epsfig}
\usepackage{amssymb}
\usepackage{float}
\usepackage{subfigure}
\usepackage{epstopdf}
\usepackage[colorlinks=true,citecolor=blue,linkcolor=blue,urlcolor=black]{hyperref}
\usepackage{dcolumn}
\usepackage{psfrag}
\usepackage{wrapfig}
\usepackage{makeidx}
\usepackage{epsf}
\usepackage{color}
\usepackage{multirow}
\usepackage{mathtools}

\begin{document}

\title{Junction conditions and thin-shells in perfect-fluid $f\left(R,T\right)$ gravity}

\author{João Luís Rosa}
\email{joaoluis92@gmail.com}
\affiliation{Institute of Physics, University of Tartu, W. Ostwaldi 1, 50411 Tartu, Estonia}

\date{\today}

\begin{abstract} 
In this work we derive the junction conditions for the matching between two spacetimes at a separation hypersurface in the perfect-fluid version of $f\left(R,T\right)$ gravity, not only in the usual geometrical representation but also in a dynamically equivalent scalar-tensor representation. We start with the general case in which a thin-shell separates the two spacetimes at the separation hypersurface, for which the general junction conditions are deduced, and the particular case for smooth matching is considered when the stress-energy tensor of the thin-shell vanishes. The set of junction conditions is similar to the one previously obtained for $f\left(R\right)$ gravity but features also constraints in the continuity of the trace of the stress-energy tensor $T_{ab}$ and its partial derivatives, which force the thin-shell to satisfy the equation of state of radiation $\sigma=2p_t$. As a consequence, a necessary and sufficient condition for spherically symmetric thin-shells to satisfy all the energy conditions is the positivity of its energy density $\sigma$. For specific forms of the function $f\left(R,T\right)$, the continuity of $R$ and $T$ ceases to be mandatory but a gravitational double-layer arises at the separation hypersurface. The Martinez thin-shell system and a thin-shell surrounding a central black-hole are provided as examples of application.
\end{abstract}


\maketitle


\section{Introduction}\label{sec:intro}

As one searches for solutions to the Einstein's field equations in General Relativity (GR) or the modified field equations in modified gravity, one might encounter a situation in which a hypersurface separates the whole spacetime into two regions described by different metric tensors expressed in two different coordinate systems. The junction conditions are the conditions the two spacetimes must satisfy in order for them to be matched at the separation hypersurface and correspond to a full solution of the field equations.

In general relativity, for the matching between the two spacetimes to be smooth one needs to guarantee that the induced metric at the separation hypersurface and the extrinsic curvature are continuous \cite{darmois,lichnerowicz,Israel:1966rt,papa,taub}. These conditions were used numerous times to derive new solutions for the Einstein's field equations in GR, e.g. constant density fluid stars with Schwarzschild exteriors, the Oppenheimer-Snyder stellar collapse \cite{oppenheimer}, and the matching between FLRW spacetimes with Vaidya exteriors \cite{senovilla1}.

The matching between the two regions can still be done even if the extrinsic curvature is not continuous, but it implies the existence of a thin-shell of matter at the separation hypersurface \cite{Israel:1966rt,lanczos1,lanczos2}. These thin-shells have been extensively studied from a thermodynamic point of view \cite{Martinez:1996ni} and the shell's entropy has been computed in numerous situations e.g. rotating \cite{Lemos:2017mci,Lemos:2017aol} and electrically charged \cite{Lemos:2015ama,Lemos:2016pyc} shells. Collisions of thin-shells have also been studied numerically \cite{brito} and, more recently, it was shown that stable extensions of the Schwarzschild fluid sphere with thin-shells provide physically relevant models for Exotic Compact Objects \cite{rosafluid}.

In the context of modified gravity, different theories of gravity will feature their own sets of junction conditions, deduced from their respective equations of motion, including not only the modified field equations but also the equations of motion of any extra fields the theory is based upon. The junction conditions have been obtained for various theories e.g. $f\left(R\right)$ gravity with \cite{Vignolo:2018eco} and without \cite{Senovilla:2013vra,Deruelle:2007pt} torsion and in the Palatini formalism\cite{Olmo:2020fri}, scalar-tensor theories \cite{Barrabes:1997kk,suffern}, Gauss-Bonnet gravity \cite{Davis:2002gn}, and more recently the hybrid metric-Palatini extension of $f\left(R\right)$ \cite{rosaworm}. The derivation of the junction conditions of these theories is not only important to allow for the development of new solutions but it is also an essential step to extend their applicability range to the rising field of exotic compact objects\cite{cardoso}.

Theories like $f\left(R\right)$\cite{clifton,nojiri,Nojiri:2017ncd}, where the action depends on an arbitrary function of the Ricci scalar $R$, have risen in a cosmological context to address the late-time cosmic acceleration period of the universe \cite{perlmutter,riess} without the necessity for dark energy sources\cite{sotiriou,felice}. These theories have also succeeded in modeling the dynamics of self-gravitating systems without requiring the presence of dark matter \cite{bohmer1,bohmer2}.  Despite these successes, extensions to the $f\left(R\right)$ theories are necessary as they also present important drawbacks. In particular, the equivalent scalar-tensor representation of the theory allows one to verify that, in order to satisfy local observational constraints, it is necessary to recur to chameleon mechanisms\cite{khoury1,khoury2}, which give rise to undesirable cosmological effects.

The $f\left(R,T\right)$ arises as a generalization of $f\left(R\right)$ gravity by allowing the action to depend on an arbitrary function of the Ricci scalar $R$ and the trace of the stress-energy tensor $T$ \cite{harko}. This theory was extensively studied in a wide range of fields, from alternatives to dark matter in galactic scales \cite{zaregonbadi}, cosmological solutions \cite{velten} including reconstruction methods \cite{houndjo1,houndjo2,jamil}, stability analysis using energy conditions \cite{alvarenga}, and in the Palatini formulation \cite{wu}, to astrophysical systems e.g. white dwarfs \cite{carvalho}, isotropic \cite{deb} and anisotropic exotic compact objects \cite{maurya, bhatti}, and atmospheric models \cite{ordines}. More recently, this theory was shown to provide relevant solutions for wormhole spacetimes \cite{sahoo,mishra,moraes,banerjee}, a context where the junction conditions of other theories were proven to be particularly useful \cite{rosaworm}. In this work, we aim not only to deduce the junction conditions of this theory to extend its applicability range, but also to provide an alternative dynamically equivalent scalar-tensor representation, which was proven useful in a wide variety of topics in other theories of gravity. In particular, the dependency of the action in two scalar quantities, namely $R$ and $T$, will be responsible for a strong similarity between the scalar-tensor representation of the $f\left(R,T\right)$ gravity and the $f\left(R,\mathcal R\right)$, see e.g. \cite{rosaworm, rosacosmo, rosabrane, rosasudden, bohmer3, bombacigno}.

This paper is organized as follows. In Sec.\ref{sec:theory} we introduce the action and equations of motion of the theory, and we introduce a dynamically equivalent scalar-tensor representation; in Sec.\ref{sec:juncgeom} we compute the junction conditions in the geometrical representation of the theory for both the matching with a thin-shell and a smooth matching; in Sec.\ref{sec:juncsca} we repeat the analysis of the junction conditions but now considering the scalar-tensor representation and we use the results to emphasize the equivalence between the two approaches; in Sec.\ref{sec:apps} we use provide three examples of application, namely to study the energy conditions of spherically symmetric thin-shells, a matching considering the well-known Martinez shell in vacuum, and a the matching of a shell surrounding a central black-hole; and finally in Sec.\ref{sec:concl} we trace our conclusions.

\subsection{Notation and assumptions}

Before proceeding, let us clearly specify the notation that will be used in the following sections, more specifically in Secs. \ref{sec:juncgeom} to \ref{sec:apps}. We define $\Sigma$ as a hypersurface that separates the spacetime $\mathcal V$ into two regions, $\mathcal V^+$ and $\mathcal V^-$. Let us consider that the metric $g_{ab}^+$, expressed in coordinates $x^a_+$, is the metric in region $\mathcal V^+$ and the metric $g_{ab}^-$, expressed in coordinates $x^a_-$, is the metric in region $\mathcal V^-$, where the latin indexes run from $0$ to $3$. Let us assume that a set of coordinates $y^\alpha$ can be defined in both sides of $\Sigma$, where greek indexes run from $0$ to $2$. The projection vectors from the 4-dimensional regions $\mathcal V^\pm$ to the 3-dimensional hypersurface $\Sigma$ are $e^a_\alpha=\partial x^a/\partial y^\alpha$. We define $n^a$ to be the unit normal vector on $\Sigma$ pointing in the direction from $\mathcal V^-$ to $\mathcal V^+$. Let $l$ denote the proper distance or time along the geodesics perpendicular to $\Sigma$ and choose $l$ to be zero at $\Sigma$, negative in the region $\mathcal V^-$, and positive in the region $\mathcal V^+$. The displacement from $\Sigma$ along the geodesics parametrized by $l$ is $dx^a=n^adl$, and $n_a=\epsilon \partial_a l$, where $\epsilon$ is either $1$ or $-1$ when $n^a$ is a spacelike or timelike vector, respectively, i.e., $n^an_a=\epsilon$. Furthermore, we will be working in the formalism of distribution functions. For any quantity $X$, we define $X=X^+\Theta\left(l\right)+X^-\Theta\left(-l\right)$, where the indexes $\pm$ indicate that the quantity $X^\pm$ is the value of the quantity $X$ in the region $\mathcal V^\pm$, and $\Theta\left(l\right)$ is the Heaviside distribution function, with $\delta\left(l\right)=\partial_l\Theta\left(l\right)$ the Dirac distribution function. We also denote $\left[X\right]=X^+|_\Sigma-X^-|_\Sigma$ as the jump of $X$ across $\Sigma$, which implies by definition that $\left[n^a\right]=\left[e^a_\alpha\right]=0$.

\section{Action and field equations}\label{sec:theory}

\subsection{Geometrical representation}

The $f\left(R,T\right)$ theory of gravity is described by an action $S$ of the form
\begin{equation}\label{action}
S=\frac{1}{2\kappa^2}\int_\Omega\sqrt{-g}f\left(R,T\right)d^4x+\int_\Omega\sqrt{-g}\mathcal L_m d^4x,
\end{equation}
where $\kappa^2\equiv \frac{8\pi G}{c^4}$, $G$ is the gravitational constant, $c$ is the speed of light, $\Omega$ is the spacetime manifold, $g$ is the determinant of the spacetime metric $g_{ab}$, $R\equiv G^{ab}R_{ab}$ is the Ricci scalar of the metric $g_{ab}$, where $R_{ab}$ is the Ricci tensor, $T\equiv g^{ab}T_{ab}$ is the trace of the stress-energy tensor $T_{ab}$, $f\left(R,T\right)$ is a well-behaved function of $R$ and $T$, and $\mathcal L_m$ is the matter Lagrangian, considered to me minimally coupled to the metric $g_{ab}$.

A variation of the action in Eq. \eqref{action} with respect to the metric $g_{ab}$ yields the modified field equations for the $f\left(R,T\right)$ theory in the form
\begin{eqnarray}
\frac{\partial f}{\partial R}R_{ab}&-&\frac{1}{2}f\left(R,T\right)g_{ab}-\left(\nabla_a\nabla_b-g_{ab}\Box\right)\frac{\partial f}{\partial R}=\nonumber \\
&=&8\pi T_{ab}-\frac{\partial f}{\partial T}\left(T_{ab}+\Theta_{ab}\right),\label{field}
\end{eqnarray}
where $\nabla_a$ is the covariant derivative and $\Box\equiv \nabla^c\nabla_c$ is the D'Alembert operator, both defined in terms of the metric $g_{ab}$, the stress-energy tensor $T_{ab}$ is defined in terms of the variation of the matter Lagrangian $\mathcal L_m$ with respect to the metric $g_{ab}$ in the usual way as
\begin{equation}
T_{ab}=-\frac{2}{\sqrt{-g}}\frac{\delta\left(\sqrt{-g}\mathcal L_m\right)}{\delta g^{ab}},
\end{equation}
and the tensor $\Theta_{ab}$ is defined in terms of the variation of the stress-energy tensor $T_{ab}$ with respect to the metric $g_{ab}$ as
\begin{equation}\label{deftheta}
\Theta_{ab}=g^{cd}\frac{\delta T_{cd}}{\delta g^{ab}}.
\end{equation}
An explicit calculation of the tensor $\Theta_{ab}$ requires a previous knowledge of the form of the stress-energy tensor $T_{ab}$ or, equivalently, of the matter Lagrangian $\mathcal L_m$. In this work, we shall assume that matter is described by an isotropic perfect fluid with an energy density $\rho$, a pressure $p$, and a four-velocity $u^a$. Under these assumptions, the stress energy tensor $T_{ab}$ takes the form
\begin{equation}\label{deftab}
T_{ab}=\left(\rho+p\right)u_au_b-pg_{ab},
\end{equation}
where the four-velocity $u^a$ satisfies the normalization property $u_au^a=1$, and for which the matter lagrangian can be written in the form $\mathcal L_m=-p$. Consequently, in this particular case Eq.\eqref{deftheta} becomes
\begin{equation}\label{thetapf}
\Theta_{ab}=-2T_{ab}-pg_{ab}.
\end{equation}

The result in Eq.\eqref{thetapf} allows us to rewrite the field equations in Eq.\eqref{field} in the more convenient form
\begin{eqnarray}
f_R R_{ab}&-&\frac{1}{2}f\left(R,T\right)g_{ab}-\left(\nabla_a\nabla_b-g_{ab}\Box\right)f_R=\nonumber \\
&=&\left(8\pi+f_T\right) T_{ab}+f_T\ p\ g_{ab},\label{field2}
\end{eqnarray}
where the subscripts $R$ and $T$ denote partial derivatives with respect to these variables. Considering that the function $f$ is a function of the two variables $R$ and $T$, then one can make use of the chain rule to write the partial derivatives $\partial_a f_X$, and the covariant derivatives $\nabla_a\nabla_bf_X$, where $X$ Represents generically $R$ or $T$, as
\begin{equation}\label{partialf}
\partial_a f_X= f_{XR}\partial_aR+f_{XT}\partial_aT,
\end{equation}
\begin{eqnarray}
&&\nabla_a\nabla_bf_X=f_{XR}\nabla_a\nabla_bR+f_{XT}\nabla_a\nabla_bT+\label{ppartialf}\\
&&+f_{XRR}\partial_aR\partial_bR+f_{XTT}\partial_aT\partial_bT+2f_{XRT}\partial_{(a}R\partial_{b)}T.\nonumber
\end{eqnarray}
The results in Eqs.\eqref{partialf} and \eqref{ppartialf}, along with the D'Alembert operator $\Box=\nabla^c\nabla_c$ which can be obtained via a contraction of Eq.\eqref{ppartialf}, allow us to expand the differential terms of the function $f$ in Eq.\eqref{field2} and write them as differential terms of both $R$ and $T$. We do not show the resultant equations explicitly due to their size.

\subsection{Equivalent scalar-tensor representation}\label{sec:scaten}

The action expressed in Eq.\eqref{action} can be recast in a dynamically equivalent scalar-tensor representation which has been proven useful in other modified theories of gravity. This can be achieved via the introduction of two auxiliary fields $\alpha$ and $\beta$ as
\begin{eqnarray}
S&=&\frac{1}{2\kappa^2}\int_\Omega\sqrt{-g}\left[f\left(\alpha,\beta\right)+\frac{\partial f}{\partial \alpha}\left(R-\alpha\right)+\right.\nonumber \\
&+&\left.\frac{\partial f}{\partial\beta}\left(T-\beta\right)\right]d^4x+\int_\Omega\sqrt{-g}\mathcal L_m d^4x\label{auxaction1}
\end{eqnarray}
The action in Eq.\eqref{auxaction1} is now a function of three independent variables, namely the metric $g_{ab}$ and the two auxiliary fields $\alpha$ and $\beta$. The equations of motion for the fields $\alpha$ and $\beta$ can be obtained from a variation of Eq.\eqref{auxaction1} with respect to these fields, respectively, and read
\begin{equation}\label{auxsys1}
f_{\alpha\alpha}\left(R-\alpha\right)+f_{\alpha\beta}\left(T-\beta\right)=0,
\end{equation}
\begin{equation}\label{auxsys2}
f_{\beta\alpha}\left(R-\alpha\right)+f_{\beta\beta}\left(T-\beta\right)=0,
\end{equation}
where we have introduced the subscript notation $\alpha$ and $\beta$ to represent partial derivatives of the function $f$ with respect to these variables, respectively. The system of Eqs.\eqref{auxsys1} and \eqref{auxsys2} can be rewritten in a matrix form $\mathcal M \textbf{x}=0$ as
\begin{equation}\label{matrixeq}
\mathcal M\textbf{x}=\begin{pmatrix}
f_{\alpha\alpha} & f_{\alpha\beta} \\
f_{\beta\alpha} & f_{\beta\beta}
\end{pmatrix}
\begin{pmatrix}
R-\alpha \\
T-\beta
\end{pmatrix}=0.
\end{equation}
Assuming that the function $f\left(\alpha,\beta\right)$ satisfies the Schwartz theorem, i.e., its crossed partial derivatives are the same, $f_{\alpha\beta}=f_{\beta\alpha}$, the solution for the system of Eqs.\eqref{auxsys1} and \eqref{auxsys2} will be unique if and only if the determinant of the matrix $\mathcal M$ does not vanish. The condition $\det\mathcal M\neq 0$ yields the relationship $f_{\alpha\alpha}f_{\beta\beta}\neq f_{\alpha\beta}^2$. Whenever this relationship is satisfied, the solution for Eqs.\eqref{auxsys1} and \eqref{auxsys2} is unique and is given by $\alpha=R$ and $\beta=T$. Inserting these results into Eq.\eqref{auxaction1} one recovers Eq.\eqref{action}, thus proving that the two representations of the theory are equivalent.

Let us now define two dynamical scalar fields $\varphi$ and $\psi$ and a scalar interaction potential $V\left(\varphi,\psi\right)$ as
\begin{equation}\label{defsca}
\varphi=\frac{\partial f}{\partial R}\qquad \psi=\frac{\partial f}{\partial T},
\end{equation}
\begin{equation}\label{defpot}
V\left(\varphi,\psi\right)=-f\left(\alpha,\beta\right)+\alpha\varphi+\beta\psi,
\end{equation}
one can rewrite the action in Eq.\eqref{auxaction1} in the equivalent scalar-tensor representation
\begin{eqnarray}
S&=&\frac{1}{2\kappa^2}\int_\Omega\sqrt{-g}\left[\varphi R+\psi T - V\left(\varphi,\psi\right)\right]d^4x\nonumber \\
&+&\int_\Omega\sqrt{-g}\mathcal L_m d^4x.\label{actionst}
\end{eqnarray}
Similarly to what happens in the metric representation of $f\left(R\right)$ theories of gravity, the scalar field $\varphi$ is analogous to a Brans-Dicke scalar field with a parameter $\omega_{BD}=0$, and with an interaction potential $V$. In addition to this scalar field, the second scalar degree of freedom of the theory associated to the arbitrary dependence of the action in $T$ is also represented by a scalar field $\psi$. 

The action in Eq.\eqref{actionst} is a function of three independent variables, the metric $g_{ab}$ and the two scalar fields $\varphi$ and $\psi$. Performing a variation of Eq.\eqref{actionst} with respect to these variables, yields the equations of motion
\begin{eqnarray}
&&\varphi R_{ab}-\frac{1}{2}g_{ab}\left(\varphi R+\psi T-V\right)- \label{fieldst}\\
&&-\left(\nabla_a\nabla_b-g_{ab}\Box\right)\varphi=8\pi T_{ab}-\psi\left(T_{ab}+\Theta_{ab}\right),\nonumber
\end{eqnarray}
\begin{equation}\label{eomphi}
V_\varphi=R,
\end{equation}
\begin{equation}\label{eompsi}
V_\psi=T,
\end{equation}
where the subscripts $\varphi$ and $\psi$ represent partial derivatives of $V$ with respect to these variables, respectively. Note that Eq.\eqref{fieldst} could be obtained directly from Eq.\eqref{field} via the direct introduction of the definitions in Eqs.\eqref{defsca} and \eqref{defpot}. 

We emphasize that in this work we shall assume that matter is described matter is well described by an isotropic perfect fluid with an energy density $\rho$ and a pressure $p$. The stress-energy tensor is again given by Eq.\eqref{deftab}. In this case, one can write the matter Lagrangian as $\mathcal L_m=-p$ and the tensor $\Theta_{ab}$ as in Eq.\eqref{thetapf}. Introducing these considerations into Eq.\eqref{fieldst}, one obtains the more convenient form of the field equations as
\begin{eqnarray}
&&\varphi R_{ab}-\frac{1}{2}g_{ab}\left(\varphi R+\psi T-V\right)-\label{fieldstpf}\\
&&-\left(\nabla_a\nabla_b-g_{ab}\Box\right)\varphi=\left(8\pi +\psi\right)T_{ab}+p\ \psi\ g_{ab}.\nonumber
\end{eqnarray}

\section{Junction conditions in the geometrical representation}\label{sec:juncgeom}

\subsection{Matching with a thin-shell at $\Sigma$}\label{sec:jcts1}

Let us now derive the junction conditions of the theory using the distribution formalism. For the spacetime $\mathcal V$ to be equipped with a metric on both sides of the hypersurface $\Sigma$, this has to be properly defined throughout the entire spacetime. In the distribution formalism, one writes the metric in the form
\begin{equation}\label{metric1}
g_{ab}=g_{ab}^+\Theta\left(l\right)+g_{ab}^-\Theta\left(-l\right).
\end{equation}
The partial derivatives of the metric defined above will take the form $\partial_cg_{ab}=\partial_c g_{ab}^+\Theta\left(l\right)+\partial_cg_{ab}^-\Theta\left(-l\right)+\epsilon\left[g_{ab}\right]n_c\delta\left(l\right)$. The apparition of a term proportional to $\delta\left(l\right)$ in these derivatives is problematic. Upon the construction of the Christoffel symbols associated to the metric $g_{ab}$ in the distribution formalism, these terms would cause products of the form $\Theta\left(l\right)\delta\left(l\right)$ to arise, which are undefined in the distribution formalism. To avoid the presence of these pathological terms, one has to impose $\left[g_{ab}\right]=0$. Furthermore, the metric $h_{\alpha\beta}$ induced on $\Sigma$ by $g_{ab}$ can be written as $h_{\alpha\beta}=g_{ab}e^a_\alpha e^b_\beta$, and consequently the induced metric from the exterior is $h_{\alpha\beta}^+=g_{ab}^+e^a_\alpha e^b_\beta$, and the induced metric from the interior is $h_{\alpha\beta}^-=g_{ab}^-e^a_\alpha e^b_\beta$. Since $\left[g_{ab}\right]=0$, we must have $h_{\alpha\beta}^+-h_{\alpha\beta}^-=0$ to preserve the continuity of the metric at $\Sigma$. We thus obtain the first junction condition as
\begin{equation}\label{junction1.1}
\left[h_{\alpha\beta}\right]=0.
\end{equation}
This junction condition corresponds also to the first junction condition in general relativity, and it does hold generically for numerous theories of gravity. Imposing Eq.\eqref{junction1.1} into the partial derivatives of Eq.\eqref{metric1}, one obtains
\begin{equation}\label{dmetric1}
\partial_cg_{ab}=\partial_c g_{ab}^+\Theta\left(l\right)+\partial_cg_{ab}^-\Theta\left(-l\right).
\end{equation}
Using Eq.\eqref{dmetric1}, one can now construct the Christoffel symbols associated with the metric $g_{ab}$ without giving rise to undefined terms, and from these one is also able to construct the Ricci tensor $R_{ab}$ and the Ricci scalar $R$. In general, $R_{ab}$ and $R$ can be written in the distribution formalism in the forms
\begin{eqnarray}\label{rab1.1}
R_{ab}&=&R_{ab}^+\Theta\left(l\right)+R_{ab}^-\Theta\left(-l\right)-\nonumber \\
&-&\left(\epsilon e_a^\alpha e_b^\beta
\left[K_{\alpha\beta}\right]+n_an_b\left[K\right]\right)\delta\left(l\right),
\end{eqnarray}
\begin{equation}\label{ricci1.1}
R=R^+\Theta\left(l\right)+R^-\Theta\left(-l\right)-2\epsilon\left[K\right]\delta\left(l\right),
\end{equation}
where $K_{\alpha\beta}=\nabla_\alpha n_\beta$ is the extrinsic curvature of the hypersurface $\Sigma$ where $n_\beta=e^b_\beta n_b$, and $K=K^\alpha_\alpha$ is the corresponding trace. The field equations in Eq.\eqref{field2} have an explicit dependence on the function $f\left(R,T\right)$ and its derivatives, which will in general feature products and power-laws of $R$ and $T$. Given the presence of terms proportional to $\delta\left(l\right)$ in the Ricci scalar $R$ (see Eq.\eqref{ricci1.1}), the function $f\left(R,T\right)$ will in general feature products of the form $\Theta\left(l\right)\delta\left(l\right)$, which are undefined in the distribution formalism, or of the form $\delta^2\left(l\right)$, which are singular. To avoid the presence of these products, one has thus to force the $\delta\left(l\right)$ term in Eq.\eqref{ricci1.1} to vanish, from which one obtains the second junction condition of the theory as
\begin{equation}\label{junction1.2}
\left[K\right]=0.
\end{equation}
Although this junction condition does not appear in general relativity, it appears to be quite common in theories of gravity where the action can be a general function of the Ricci scalar $R$, like $f\left(R\right)$ or hybrid metric-Palatini gravity. The forms of the Ricci tensor $R_{ab}$ and the Ricci scalar $R$ from Eqs.\eqref{rab1.1} and \eqref{ricci1.1} thus simplify to
\begin{equation}\label{rab1.2}
R_{ab}=R_{ab}^+\Theta\left(l\right)+R_{ab}^-\Theta\left(-l\right)-\epsilon e_a^\alpha e_b^\beta\left[K_{\alpha\beta}\right]\delta\left(l\right),
\end{equation}
\begin{equation}\label{ricci1.2}
R=R^+\Theta\left(l\right)+R^-\Theta\left(-l\right),
\end{equation}
respectively. 

Given the presence of differential terms $\nabla_a\nabla_bf_R$ and $\Box f_R$ in the field equations in Eq.\eqref{field2}, one must also look into the first and second order derivatives of the Ricci scalar $R$. Taking the partial derivative of Eq.\eqref{ricci1.2} one obtains
\begin{equation}\label{dricci1.1}
\partial_a R=\partial_aR^+\Theta\left(l\right)+\partial_aR^-\Theta\left(-l\right)+\epsilon\left[R\right]n_a\delta\left(l\right).
\end{equation}
According to the expansion of the differential terms shown in Eq.\eqref{ppartialf}, the field equations will feature products of partial derivatives of $R$, i.e., terms of the form $\partial_aR\partial_bR$. Due to the existence of terms proportional to $\delta\left(l\right)$ in $\partial_aR$, as can be seen in Eq.\eqref{dricci1.1}, the terms $\partial_aR\partial_bR$ will also feature products of the form $\Theta\left(l\right)\delta\left(l\right)$ and $\delta^2\left(l\right)$, which are undefined and singular in the distribution formalism, respectively. Thus, to avoid the presence of these terms, one has to impose a third junction condition of the form
\begin{equation}\label{junction1.3}
\left[R\right]=0.
\end{equation}
This junction condition is also not present in general relativity, but it does appear in modified theories of gravity with an extra dynamical scalar degree of freedom associated with a function of the Ricci scalar, e.g. the metric formulation of $f\left(R\right)$. Eq.\eqref{junction1.3} allow us to simplify Eq.\eqref{dricci1.1} to
\begin{equation}\label{dricci1.2}
\partial_a R=\partial_aR^+\Theta\left(l\right)+\partial_aR^-\Theta\left(-l\right),
\end{equation}
and finally we are able to compute the second-order covariant derivatives of the Ricci scalar $\nabla_a\nabla_b R$, which take the general form
\begin{eqnarray}
\nabla_a\nabla_b{R}&=&\nabla_a\nabla_b{R}_+\Theta\left(l\right)+\nabla_a\nabla_b{R}_-\Theta\left(-l\right)+\nonumber \\
&+&\epsilon n_a\left[\partial_b{R}\right]\delta\left(l\right).\label{ddricci1.1}
\end{eqnarray}

Let us now turn to the matter sector of the theory. In the previous paragraphs, more specifically in Eqs.\eqref{rab1.2} and \eqref{ddricci1.1}, we have shown that the Ricci tensor $R_{ab}$ and the second order derivatives $\nabla_a\nabla_b R$ feature terms proportional to $\delta\left(l\right)$, which will consequently be present in the left-hand side of the field equations in Eq.\eqref{field2}. These terms can be associated with the presence of a thin-shell of matter at the separation hypersurface $\Sigma$. To find the properties of this thin-shell, let us write the stress-energy tensor $T_{ab}$ as a distribution function of the form
\begin{equation}\label{set1.1}
T_{ab}=T_{ab}^+\Theta\left(l\right)+T_{ab}^-\Theta\left(-l\right)+\delta\left(l\right)S_{ab},
\end{equation} 
where $S_{ab}$ is the four-dimensional stress-energy tensor of the thin-shell, which can be written as a three-dimensional tensor at $\Sigma$ as
\begin{equation}
S_{ab}=S_{\alpha\beta}e^\alpha_ae^\beta_b.
\end{equation}
Taking the trace of Eq.\eqref{set1.1} one finds that the trace $T$ of the stress-energy tensor also features a ter proportional to $\delta\left(l\right)$, as $T=T^+\Theta\left(l\right)+T^-\Theta\left(-l\right)+\delta\left(l\right)S$, where $S=S^a_a$ is the trace of the stress-energy tensor of the thin-shell. Following the same argument that lead to Eq.\eqref{junction1.2}, i.e., the fact that the field equations will in general depend on products and power-laws of $T$ through the function $f\left(R,T\right)$ and its partial derivatives, we conclude that this function will in general depend on products of the forms $\Theta\left(l\right)\delta\left(l\right)$ and $\delta^2\left(l\right)$ which are undefined or singular in the distribution formalism. To avoid the presence of these products, one has thus to force the trace of the stress-energy tensor of the thin-shell to vanish, and we obtain the fourth junction condition
\begin{equation}\label{junction1.4}
S=0.
\end{equation} 
This junction condition also does not appear in general relativity and it does not appear in other well-known theories as $f\left(R\right)$, at is is associated with an extra scalar degree of freedom associated with an arbitrary dependence of the action in $T$. Eq.\eqref{junction1.4} allows us to write the trace of the stress-energy tensor $T$ in the simplified form
\begin{equation}\label{T1.1}
T=T^+\Theta\left(l\right)+T^-\Theta\left(-l\right).
\end{equation}
The presence of the differential terms $\Box f_R$ and $\nabla_a\nabla_b f_R$ in the field equations in Eq.\eqref{field2} imply that one must also take into account the first and second covariant derivatives of $T$ in the distribution formalism. Taking the partial derivative of Eq.\eqref{T1.1}, one obtains
\begin{equation}\label{dT1.1}
\partial_a T=\partial_aT^+\Theta\left(l\right)+\partial_aT^-\Theta\left(-l\right)+\epsilon\left[T\right]n_a\delta\left(l\right).
\end{equation}
Provided the expansion of the differential terms in Eq.\eqref{ppartialf}, one verifies that the field equations in Eq.\eqref{field2} will feature products of the partial derivatives of $T$ in the form $\partial_a T\partial_b T$. Due to the presence of terms proportional to $\delta\left(l\right)$ in Eq.\eqref{dT1.1}, the terms $\partial_a T\partial_b T$ will feature products of the forms $\Theta\left(l\right)\delta\left(l\right)$ and $\delta^2\left(l\right)$ which are undefined or singular in the distribution formalism. To avoid the presence of these terms, one has to impose a fifth junction condition as
\begin{equation}\label{junction1.5}
\left[T\right]=0.
\end{equation}
Again, this junction condition does not appear in general relativity of $f\left(R\right)$ theories. Eq.\eqref{junction1.5} allow us to rewrite Eq.\eqref{dT1.1} in the simplified form
\begin{equation}\label{dT1.2}
\partial_a T=\partial_aT^+\Theta\left(l\right)+\partial_aT^-\Theta\left(-l\right).
\end{equation}
Finally, we are able to compute the second-order covariant derivatives of the trace $T$, $\nabla_a\nabla_b T$, which can be written in the distribution formalism as
\begin{eqnarray}
\nabla_a\nabla_bT&=&\nabla_a\nabla_bT_+\Theta\left(l\right)+\nabla_a\nabla_bT_-\Theta\left(-l\right)+\nonumber \\
&+&\epsilon n_a\left[\partial_bT\right]\delta\left(l\right).\label{ddT1.1}
\end{eqnarray}

We are now in conditions of determining the stress-energy tensor $S_{\alpha\beta}$ of the thin-shell. To do so, one introduces the distribution formalism representations given in Eqs.\eqref{metric1}, \eqref{rab1.2}, \eqref{ricci1.2}, \eqref{dricci1.2}, \eqref{ddricci1.1}, \eqref{set1.1}, \eqref{T1.1}, \eqref{dT1.2}, and \eqref{ddT1.1} into the field equations in Eq.\eqref{field2} and project the result into the hypersurface $\Sigma$ using $e^a_\alpha e^b_\beta$. The result is as follows
\begin{eqnarray}
\left(8\pi+f_T\right)S_{\alpha\beta}=-\epsilon f_R\left[K_{\alpha\beta}\right]+\nonumber \\
+\epsilon h_{\alpha\beta}n^c\left(f_{RR}\left[\partial_cR\right]+f_{RT}\left[\partial_cT\right]\right).\label{sab1.1}
\end{eqnarray}
Taking the trace of Eq.\eqref{sab1.1}, inserting the result into Eq.\eqref{junction1.4}, and using the second junction condition given in Eq.\eqref{junction1.2}, allows us to rewrite the fourth junction condition in the more convenient form $n^c\left(f_{RR}\left[\partial_cR\right]+f_{RT}\left[\partial_cT\right]\right)=0$, from which we realize that the second term on the right-hand side of Eq.\eqref{sab1.1} must vanish. 

To summarize, the complete set of junction conditions for the $f\left(R,T\right)$ gravity in the general case of a matching with a thin-shell at $\Sigma$ is thus composed of the following six equations
\begin{eqnarray}
&\left[h_{\alpha\beta}\right]=0,\nonumber \\
&\left[K\right]=0,\nonumber \\
&\left[R\right]=0, \label{fullset1}\\
&\left[T\right]=0,\nonumber\\
&n^c\left(f_{RR}\left[\partial_cR\right]+f_{RT}\left[\partial_cT\right]\right)=0, \nonumber \\
&\left(8\pi+f_T\right)S_{\alpha\beta}=-\epsilon f_R\left[K_{\alpha\beta}\right]. \nonumber
\end{eqnarray}

\subsection{Smooth matching at $\Sigma$}

In the previous section, we have considered a matching between two spacetime regions $\mathcal V^\pm$ at a separation hypersurface $\Sigma$ with the presence of a matter thin-shell at $\Sigma$ described by a tress-energy tensor $S_{\alpha\beta}$. In the particular case where $S_{\alpha\beta}$ vanishes, the matching between the two spacetime regions is smooth,  i.e., no thin-shell is needed at $\Sigma$. A different set of junction conditions can be obtained in this particular case. As the presence of the thin-shell is associated with the terms proportional to $\delta\left(l\right)$ in the field equations in Eq.\eqref{field2}, the smooth matching case can be obtained by forcing these terms to vanish. We now pursue such analysis.

Let us start again with the metric $g_{ab}$. The form of the metric provided in Eq.\eqref{metric1} in the distribution formalism is the same for the smooth matching, as it does not depend on $\delta\left(l\right)$. Following the same reasoning as in Sec.\ref{sec:jcts1}, we conclude that the induced metric $h_{\alpha\beta}$ at $\Sigma$ must be continuous, and the first junction condition reads
\begin{equation}\label{junction2.1}
\left[h_{\alpha\beta}\right]=0.
\end{equation}
Similarly as before, one can now compute the Christoffel symbols, the Ricci Tensor $R_{ab}$ and the Ricci scalar $R$ associated with this metric. The Ricci tensor and the Ricci scalar will still have the same general forms as given in Eqs.\eqref{rab1.1} and \eqref{ricci1.1}, respectively, where $K_{\alpha\beta}=\nabla_\alpha n_\beta$ is the extrinsic curvature of the hypersurface $\Sigma$. From Eq.\eqref{field2}, one verifies that the field equations have a term proportional to $R_{ab}$, which according to Eq.\eqref{rab1.1} presents terms proportional to $\delta\left(l\right)$. As these terms must not be present for the matching to be smooth, one must impose that the extrinsic curvature $K_{\alpha\beta}$ is continuous at $\Sigma$, i.e., we obtain the second junction condition
\begin{equation}\label{junction2.2}
\left[K_{\alpha\beta}\right]=0.
\end{equation}
This junction condition also appears in general relativity for the particular case of smooth matching and, along with Eq.\eqref{junction2.2}, is a common junction condition to appear in metric theories of gravity. As Eq.\eqref{junction2.2} imposes directly that $\left[K\right]=0$, this latter equations does not have to be imposed separately. The Ricci tensor $T_{ab}$ and the Ricci scalar $R$ thus take the forms
\begin{equation}\label{rab2.2}
R_{ab}=R_{ab}^+\Theta\left(l\right)+R_{ab}^-\Theta\left(-l\right),
\end{equation}
\begin{equation}\label{ricci2.2}
R=R^+\Theta\left(l\right)+R^-\Theta\left(-l\right).
\end{equation}
Note that the resultant form of $R$ in Eq.\eqref{ricci2.2} already guarantees that the products and power-laws in $R$ arising from the terms proportional to $f\left(R,T\right)$ and its partial derivatives in the field equations in Eq.\eqref{field2} are regular.

Turning now to the differential terms $\nabla_a\nabla_b f_R$ and $\Box f_R$ in Eq.\eqref{field2}, we have to analyze the first and second-order covariant derivatives or the Ricci scalar $R$. As the form of $R$ in Eq.\eqref{ricci2.2} is the same as in Eq.\eqref{ricci1.2}, the partial derivatives of $R$ are still of the form given in Eq.\eqref{dricci1.1}. Thus, following the same arguments as in Sec.\ref{sec:jcts1}, we conclude that the Ricci scalar must be continuous across $\Sigma$, and the third junction condition becomes
\begin{equation}\label{junction2.3}
\left[R\right]=0.
\end{equation}
Given this condition, one is again able to write the partial derivatives $\partial_a R$ to the simplified form in Eq.\eqref{dricci1.2}. Taking the covariant derivative of Eq.\eqref{dricci1.2}, one again obtains the second order derivatives $\nabla_a\nabla_b R$ in the form of Eq.\eqref{ddricci1.1}. In Sec.\ref{sec:jcts1}, the presence of the terms proportional to $\delta\left(l\right)$ in $\nabla_a\nabla_b$ were not pathological as we admitted the existence of a thin-shell of matter at $\Sigma$. However, since these terms are present in the field equations in Eq.\eqref{field2} through the differential terms (see Eq.\eqref{ppartialf}), one has to force the continuity of $\partial_a R$. We thus obtain the fourth junction condition for smooth matching as
\begin{equation}\label{junction2.4}
\left[\partial_cR\right]=0.
\end{equation} 
This junction condition appears commonly for the smooth matching in theories of gravity with a scalar degree of freedom associated with an arbitrary function of $R$ in the action, e.g. in $f\left(R\right)$ and hybrid metric-Palatini. Eq.\eqref{junction2.4} allows us to write $\nabla_a\nabla_b R$ in the simplified form
\begin{equation}
\nabla_a\nabla_b{R}=\nabla_a\nabla_b{R}_+\Theta\left(l\right)+\nabla_a\nabla_b{R}_-\Theta\left(-l\right).
\end{equation}

Let us now turn to the matter sector of the theory. For the matching between the two spacetime regions to be smooth, one must avoid the presence of terms proportional to $\delta\left(l\right)$ in the field equations in Eq.\eqref{field2}. Thus, the stress-energy tensor $T_{ab}$ and its trace $T$ must be forcefully of the forms
\begin{equation}\label{set2}
T_{ab}=T_{ab}^+\Theta\left(l\right)+T_{ab}^-\Theta\left(-l\right),
\end{equation}
\begin{equation}\label{T2}
T=T^+\Theta\left(l\right)+T^-\Theta\left(-l\right).
\end{equation}
Note that the form of $T$ in Eq.\eqref{T2} guarantees that the products and power-laws of $T$ arising in the function $f\left(R,T\right)$ and its derivatives are regular. Similarly as before, the presence of the differential terms $\nabla_a\nabla_b f_R$ and $\Box f_R$ in Eq.\eqref{field2} imply that we also have to analyze the first and second derivatives of the trace of the stress-energy tensor $T$. Since the form of $T$ in Eq.\eqref{T2} is the same as in Eq.\eqref{T1.1}, the partial derivatives $\partial_a T$ will also be of the same form as in Eq.\eqref{dT1.1}. Thus, following the same arguments as in Sec.\ref{sec:jcts1}, we conclude that $T$ must be continuous across $\Sigma$, and the fifth junction condition becomes
\begin{equation}\label{junction2.5}
\left[T\right]=0.
\end{equation}
Under Eq.\eqref{junction2.5} one is again able to write $\partial_a T$ in the form of Eq.\eqref{dT1.2}. Finally, taking a covariant derivative of Eq.\eqref{dT1.2}, we obtain the second order derivatives $\nabla_a\nabla_b T$ in the same form of Eq.\eqref{ddT1.1}. Similarly to what happens for $\nabla_a\nabla_b R$, in the case of a smooth matching at $\Sigma$ the terms proportional to $\delta\left(l\right)$ in Eq.\eqref{ddT1.1} must vanish, as otherwise we would have terms proportional to $\delta\left(l\right)$ in the field equations due to the expansion in Eq.\eqref{ppartialf}. Thus, one must impose continuity in $\partial_a T$, and we obtain the sixth junction condition in the form
\begin{equation}\label{junction2.6}
\left[\partial_c T\right]=0.
\end{equation}

To summarize, the complete set of junction conditions for the $f\left(R,T\right)$ gravity in the particular case of a smooth matching at $\Sigma$ is composed of the following six equations
\begin{eqnarray}
&\left[h_{\alpha\beta}\right]=0.\nonumber \\
&\left[K_{\alpha\beta}\right]=0.\nonumber \\
&\left[R\right]=0.\label{fullset2}\\
&\left[T\right]=0.\nonumber \\
&\left[\partial_c R\right]=0.\nonumber \\
&\left[\partial_c T\right]=0.\nonumber
\end{eqnarray}

\subsection{Double gravitational layers at $\Sigma$}

\subsubsection{Matching with $\left[R\right]\neq0$ and $\left[T\right]\neq0$}\label{sec:geodl1}

In Sec.\ref{sec:jcts1}, we have derived the general set of junction conditions that must be satisfied in order to match two spacetimes $\mathcal V^\pm$ at a given separation hypersurface $\Sigma$ for a general function $f\left(R,T\right)$. In particular, the junction conditions given in Eqs.\eqref{junction1.3} and \eqref{junction1.5} were imposed to avoid the presence of products of the form $\Theta\left(l\right)\delta\left(l\right)$ and $\delta^2\left(l\right)$ in the field equations, arising from the expansions of the differential terms of $f$ in Eq.\eqref{ppartialf}. 

There is, however, an alternative way of avoiding the presence of these terms without having to impose Eqs.\eqref{junction1.3} and \eqref{junction1.5}, which is the selection of particular forms of the function $f\left(R,T\right)$ for which the products $\partial_a R\partial_b R$, $\partial_a T\partial_b T$, and $\partial_a T\partial_b R$ are not present in the field equations. As can be seen from Eq.\eqref{ppartialf} with $f_X=f_R$, these products can be removed from the field equations via the choice of a function $f\left(R,T\right)$ which satisfies the conditions $f_{RRR}=f_{RRT}=f_{RTT}=0$. The most general function that satisfies these properties is given by
\begin{equation}\label{functiondl}
f\left(R,T\right)=R\left(1+\gamma T\right)-2\Lambda+\alpha R^2+g\left(T\right),
\end{equation}
where $\gamma$, $\Lambda$ and $\alpha$ are arbitrary constants and $g\left(T\right)$ is a well-behaved function of the trace of the stress-energy tensor $T$. Inserting Eq.\eqref{functiondl} into the field equations in Eq.\eqref{field2} yields the field equations for this particular case as
\begin{eqnarray}
&&\left(1+\gamma T+2\alpha R\right)R_{ab}-\left(\nabla_a\nabla_b-g_{ab}\Box\right)\left(2\alpha R+\gamma T\right)-\nonumber \\
&&-\frac{1}{2}g_{ab}\left[R\left(1+\gamma T\right)-2\Lambda+\alpha R^2+g\left(T\right)\right]=8\pi T_{ab}+\nonumber \\
&&+\left[\gamma R+ g'\left(T\right)\right]\left(T_{ab}+p\ g_{ab}\right).\label{dlfield}
\end{eqnarray}

At this point it is important to state a few remarks regarding the parameters $\alpha$, $\gamma$ and $\Lambda$. As can be seen from Eq.\eqref{functiondl}, the coefficient of $R$ in the function $f\left(R,T\right)$ is effectively $\left(1+\gamma T\right)$. In order to preserve the positivity of the Einstein-Hilbert term in the action, one must impose that $\gamma T>-1$, and thus the relevant parameter space for $\gamma$ will have to be verified case by case depending on the form of $T$. Furthermore, although there are no constraints on the sign of the quadratic term, positive values for $\alpha$ have been shown to provide useful results in the context of cosmology. In particular, the Starobinski inflation model can be attained with $\alpha=1/m^2$, where $m$ is a constant with units of mass. Finally, the convenient definition of $\Lambda$ implies that this parameter plays the role of a cosmological constant and mainly controls the asymptotics of the models, which will be asymptotically de-Sitter ($\Lambda>0$), anti de-Sitter ($\Lambda<0$), or Minkowski ($\Lambda=0$).

Let us now analyze the consequences of choosing a function $f\left(R,T\right)$ in the form of Eq.\eqref{functiondl}. In this case, the metric $g_{ab}$ and its partial derivatives $\partial_a g_{ab}$ are still given by the forms of Eqs.\eqref{metric1} and \eqref{dmetric1}, respectively, the Ricci tensor $R_{ab}$ is given by Eq.\eqref{rab1.2}, the Ricci scalar $R$ is given by Eq.\eqref{ricci1.2}, and the partial derivatives of the Ricci scalar $\partial_a R$ are given by Eq.\eqref{dricci1.1}. Thus, the analysis that leads to the junction conditions in Eqs.\eqref{junction1.1} and \eqref{junction1.2} is the same, i.e., the first and second junction conditions remain as
\begin{equation}\label{junctiondl1}
\left[h_{\alpha\beta}\right]=0,
\end{equation} 
\begin{equation}\label{junctiondl2}
\left[K\right]=0,
\end{equation}

In the matter sector, since the function $f\left(R,T\right)$ given in Eq.\eqref{functiondl} depends on an arbitrary function $g\left(T\right)$, which in general can depend on power-laws of $T$, one concludes that $T$ must not have any dependence on $\delta\left(l\right)$, as otherwise undefined products $\Theta\left(l\right)\delta\left(l\right)$ and singular products $\delta\left(l\right)^2$ would appear in the field equations in Eq.\eqref{dlfield}. Thus, the trace $T$ and its partial derivative $\partial_a T$ are given by Eqs.\eqref{T1.1} and \eqref{dT1.1}, respectively, and whichever terms proportional to $\delta\left(l\right)$ arise in $T_{ab}$ must vanish upon tracing.

The differences in comparison with the general case of Sec.\ref{sec:jcts1} arise only at the level of the second-order derivatives of $R$ and $T$, i.e, the terms $\nabla_a\nabla_b R$ and $\nabla_a\nabla_b T$, which now take the forms
\begin{equation}\label{dlddR}
\nabla_a\nabla_b R=\left(\nabla^2R\right)_{ab}+\epsilon\nabla_a\left(\left[R\right]\delta\left(l\right)n_b\right),
\end{equation}
\begin{equation}\label{dlddT}
\nabla_a\nabla_b T=\left(\nabla^2T\right)_{ab}+\epsilon\nabla_a\left(\left[T\right]\delta\left(l\right)n_b\right),
\end{equation}
where $\left(\nabla^2R\right)_{ab}$ and $\left(\nabla^2T\right)_{ab}$ collectively denote the right-hand sides of Eqs.\eqref{ddricci1.1} and \eqref{ddT1.1}, respectively. Thus, although the junction conditions $\left[R\right]=0$ and $\left[T\right]=0$ can be discarded in this particular situation, as a consequence new terms will arise in the stress-energy tensor $S_{\alpha\beta}$ of the thin-shell. The second terms on the right-hand sides of Eqs.\eqref{dlddR} and \eqref{dlddT} are also present in particular forms of $f\left(R\right)$ and hybrid metric-Palatini theories of gravity, and can be rewritten as
\begin{eqnarray}
&&\nabla_a\left(\left[R\right]\delta\left(l\right)n_b\right)=\Delta_{ab}^R+\nonumber \\
&&+\delta\left(l\right)\left(K_{ab}-\epsilon K n_an_b+n_bh_a^c\nabla_c\right)\left[R\right],\label{dlddR2}
\end{eqnarray}
\begin{eqnarray}
&&\nabla_a\left(\left[T\right]\delta\left(l\right)n_b\right)=\Delta_{ab}^T+\nonumber \\
&&+\delta\left(l\right)\left(K_{ab}-\epsilon K n_an_b+n_bh_a^c\nabla_c\right)\left[T\right],\label{dlddT2}
\end{eqnarray}
where the distribution functions $\Delta_{ab}^R$ and $\Delta_{ab}^T$ are defined as
\begin{equation}
\int_\Omega\Delta_{ab}^RY^{ab}d^4x=-\epsilon\int_\Sigma\left[R\right]n_a n_bn^c\nabla_cY^{ab}d^3x,
\end{equation}
\begin{equation}
\int_\Omega\Delta_{ab}^TY^{ab}d^4x=-\epsilon\int_\Sigma\left[T\right]n_a n_bn^c\nabla_cY^{ab}d^3x,
\end{equation}
for a given test function $Y^{ab}$. Inserting Eqs.\eqref{dlddR2} and \eqref{ddricci1.1} into Eq.\eqref{dlddR}, inserting Eq.\eqref{dlddT2} and \eqref{ddT1.1} into Eq.\eqref{dlddT}, and finally inserting the results into the field equations in Eq.\eqref{dlfield}, one verifies that in this case the stress-energy tensor $T_{ab}$ must be written in the form
\begin{eqnarray}
T_{ab}&=&T_{ab}^+\Theta\left(l\right)+T_{ab}^-\Theta\left(-l\right)+\label{dlTab}\\
&+&\delta\left(l\right)\left(S_{ab}+2S_{(a}n_{b)}+Sn_an_b\right)+s_{ab}\left(l\right),\nonumber
\end{eqnarray}
where $S_{ab}$ is the stress-energy tensor of the thin-shell, $S_a$ is the external flux momentum whose normal component measures the normal energy flux across $\Sigma$ and spacial components measure tangential stresses, $S$ is the external normal pressure or tension supported on $\Sigma$, and $s_{ab}$ is the double-layer stress-energy tensor distribution. These variables can be written in terms of the geometrical quantities in the forms
\begin{eqnarray}
&&\kappa^2_{\text{eff}}S_{ab}=-\left(1+\gamma T^\Sigma+2\alpha R^\Sigma \right)\epsilon\left[K_{ab}\right]+\label{dlsab}\\
&&+\epsilon h_{ab}n^c\left(2\alpha\left[\nabla_cR\right]+\gamma\left[\nabla_cT\right]\right)-\epsilon K_{ab}^\Sigma\left(2\alpha\left[R\right]+\gamma\left[T\right]\right),\nonumber 
\end{eqnarray}
\begin{equation}\label{dlsa}
\kappa^2_{\text{eff}}S_a=-\epsilon h^c_a\nabla_c\left(2\alpha\left[R\right]+\gamma\left[T\right]\right),
\end{equation}
\begin{equation}\label{dls}
\kappa^2_{\text{eff}}S= K^\Sigma\left(2\alpha\left[R\right]+\gamma\left[T\right]\right),
\end{equation}
\begin{equation}\label{dldsab}
\kappa^2_{\text{eff}}s_{ab}\left(l\right)=2\alpha\epsilon\Omega^R\left(l\right)+\gamma\epsilon\Omega^T\left(l\right),
\end{equation}
where we have defined $\kappa^2_{\text{eff}}=8\pi+\gamma R^\Sigma+g'\left(T^\Sigma\right)$, we have defined $R^\Sigma$, $T^\Sigma$ and $K_{ab}^\Sigma$ as the average values of these quantities at the hypersurface $\Sigma$, i.e., $2R^\Sigma=R^++R^-$ and similarly to the other quantities. The distributions $\Omega_{ab}^R\left(l\right)$ and $\Omega_{ab}^T\left(l\right)$ in Eq.\eqref{dldsab} are defined as $\Omega_{ab}^X=h_{ab}\Delta^X-\Delta^X_{ab}$, where $X$ denotes either $R$ or $T$ and $\Delta^X$ is the trace of $\Delta^X_{ab}$. According to these definitions, the double-layer stress-energy tensor distribution can be written explicitly as
\begin{equation}\label{dlsabl}
\int_\Omega \kappa^2_{\text{eff}} s_{ab}Y^{ab}d^4x=-\int_\Sigma \epsilon h_{ab}\left(2\alpha\left[R\right]+\gamma\left[T\right]\right)n^c\nabla_cY^{ab}d^3x.
\end{equation}

To preserve the regularity of the function $g\left(T\right)$ in Eq.\eqref{functiondl} we had to force the trace of the stress-energy tensor $T$ to be written as in Eq.\eqref{T1.1}. Let us now check which conditions arise from imposing this restriction into Eq.\eqref{dlTab}. Taking the trace of Eq.\eqref{dlTab}, using Eqs.\eqref{junctiondl2}, \eqref{dlsab}, \eqref{dlsa}, and \eqref{dls}, and noting that $n^ah_a^b=0$, one obtains
\begin{eqnarray}\label{Tdl}
T&=&T_{ab}^+\Theta\left(l\right)+T_{ab}^-\Theta\left(-l\right)+\nonumber\\
&+&3\epsilon\delta\left(l\right)n^c\left(2\alpha\left[\nabla_cR\right]+\gamma\left[\nabla_cT\right]\right)+s_a^a\left(l\right).
\end{eqnarray}
A comparison between Eqs.\eqref{Tdl} and \eqref{T1.1} reveals that both the terms proportional to $\delta\left(l\right)$ and the distribution $s_a^a\left(l\right)$ must vanish identically to preserve the regularity of the function $g\left(T\right)$. Consequently, one deduces the third and fourth junction conditions for this particular case as
\begin{equation}\label{junctiondl3}
2\alpha\left[\nabla_cR\right]+\gamma\left[\nabla_cT\right]=0,
\end{equation}
\begin{equation}\label{junctiondl4}
2\alpha\left[R\right]+\gamma\left[T\right]=0,
\end{equation}
where Eq.\eqref{junctiondl3} corresponds to fifth equation in the system of Eq.\eqref{fullset1} in the particular case for which the function $f\left(R,T\right)$ is given by the explicit form provided in Eq.\eqref{functiondl}. Inserting Eqs.\eqref{junctiondl3} and \eqref{junctiondl4} into Eqs. \eqref{dlsab}, \eqref{dlsa}, \eqref{dls}, and Eq.\eqref{dlsabl}, one verifies that $S_a$, $S$ and $s_{ab}\left(l\right)$ vanish completely and $S_{ab}$ reduces to the term proportional do the jump of the extrinsic curvature.

To summarize, the full set of junction conditions in the geometrical representation of the $f\left(R,T\right)$ for the particular case where the function is given by the explicit form of Eq.\eqref{functiondl} is composed of the following five equations
\begin{eqnarray}
&\left[h_{\alpha\beta}\right]=0,\nonumber\\
&\left[K\right]=0,\nonumber\\
&2\alpha\left[R\right]+\gamma\left[T\right]=0,\label{fullsetdl1}\\
&2\alpha\left[\nabla_cR\right]+\gamma\left[\nabla_cT\right]=0,\nonumber\\
&\left[8\pi+\gamma R^\Sigma+g'\left(T^\Sigma\right)\right]S_{ab}=-\left(1+\gamma T^\Sigma+2\alpha R^\Sigma \right)\epsilon\left[K_{ab}\right].\nonumber
\end{eqnarray}
Similarly to what happens in $f\left(R\right)$ gravity and other similar theories, it is possible to consider particular forms of the function $f\left(R,T\right)$ for which some of the junction conditions, namely $\left[R\right]=0$ and $\left[T\right]=0$, can be discarded from the final set of equations. However, as one needs to preserve the regularity of products and power-laws of $T$, this situation does not give rise to a gravitational double-layer at $\Sigma$. Instead, an extra junction condition arises constraining $\left[R\right]$ to be proportional to $\left[T\right]$. Forcing $\left[R\right]=0$ and $\left[T\right]=0$ in Eqs.\eqref{fullsetdl1} one recovers the general system of Eqs\eqref{fullset1} in the particular case of Eq.\eqref{functiondl}, as expected.

\subsubsection{Matching with a double-layer at $\Sigma$}\label{sec:geodl2}

In the previous example, we were able to discard the junction conditions $\left[R\right]=0$ and $\left[T\right]=0$ from the full set of junction conditions in Eq.\eqref{fullset1} via the choice of a particular form of the function $f\left(R,T\right)$. Even though the form chosen in Eq.\eqref{functiondl} is the most general form of $f\left(R,T\right)$ that allows for this simplification to happen, it is still not enough to raise the appearance of a gravitational double-layer at $\Sigma$, as the regularity of the products $RT$ and possible power-laws of $T$ in $g\left(T\right)$ force the double-layer terms to vanish. Thus, for the gravitational double-layer to appear, one needs to constraint further the function $f\left(R,T\right)$ in such a way that the products $RT$ do not appear and $g\left(T\right)$ is at most linear in $T$, i.e., we now consider
\begin{equation}\label{functiondl2}
f\left(R,T\right)=R-2\Lambda+\alpha R^2+\gamma T,
\end{equation}
where $\Lambda$, $\alpha$, and $\gamma$ are arbitrary constants. Inserting Eq.\eqref{functiondl2} into the field equations in Eq.\eqref{field2} yields the field equations for this particular case
\begin{eqnarray}
\left(1+2\alpha R\right)R_{ab}-\frac{1}{2}g_{ab}\left(R-2\Lambda+\alpha R^2+\gamma T\right)-\nonumber \\
-2\alpha\left(\nabla_a\nabla_b-g_{ab}\Box\right)R=\left(8\pi+\gamma\right)T_{ab}+\gamma\ p\ g_{ab}.\label{fielddl2}
\end{eqnarray}

Similarly to the previous case of Sec.\ref{sec:geodl1}, in this case the metric $g_{ab}$ and its partial derivatives $\partial_a g_{ab}$ are again given by the forms in Eqs.\eqref{metric1} and \eqref{dmetric1}, respectively, the Ricci tensor $R_{ab}$ is given by Eq.\eqref{rab1.2}, the Ricci scalar $R$ is given by Eq.\eqref{ricci1.2}, and the partial derivatives of the Ricci scalar $\partial_a R$ are given by Eq.\eqref{dricci1.1}. Following the same analysis as in the previous sections, one arives to the first and second junction conditions as
\begin{equation}\label{junctiondl2.1}
\left[h_{\alpha\beta}\right]=0,
\end{equation} 
\begin{equation}\label{junctiondl2.2}
\left[K\right]=0.
\end{equation}

The main difference between this case and the one in Sec.\ref{sec:geodl1} is that, since we have chosen $f\left(R,T\right)$ to be linear in $T$, the second-order differential terms in $f_R$ are now independent of $T$. Therefore, we only have to take into consideration the second-order derivatives of $R$, which take the form of Eq.\eqref{dlddR} and lead to Eqs.\eqref{dlddR2}. Inserting these results into the field equation in Eq.\eqref{fielddl2} one again verifies that the stress-energy tensor $T_{ab}$ must be written in the form of Eq.\eqref{dlTab}. In this case however, the field equations depend explicitly in the trace of the stress-energy tensor $T$ which is not anymore restricted to have vanishing contributions from $\delta\left(l\right)$ and $s_{ab}\left(l\right)$. Tracing Eq.\eqref{dlTab} one finds
\begin{equation}
T=T^+\Theta\left(l\right)+T^-\Theta\left(-l\right)+\delta\left(l\right)\left(S_a^a+\epsilon S\right)+s_a^a\left(l\right).
\end{equation}
where we have used the result $n^aS_a=0$. Inserting these considerations into the field equations, one verifies that all the independent projections of the singular part of the stress-energy tensor $T_{ab}$ in the distribution formalism, i.e., the quantities $S_{ab}$, $S_a$ and $S$, and the distribution $s_{ab}\left(l\right)$, are now given by
\begin{eqnarray}
&&\left(8\pi+\gamma\right)S_{ab}+\frac{1}{2}h_{ab}\gamma\left(S_c^c+\epsilon S\right)=\label{dlsab2} \\
&&-\left(1+2\alpha R^\Sigma \right)\epsilon\left[K_{ab}\right]+2\alpha\epsilon h_{ab}n^c\left[\nabla_cR\right]-2\alpha\epsilon K_{ab}^\Sigma\left[R\right],\nonumber
\end{eqnarray}
\begin{equation}\label{dlsa2}
\left(8\pi+\gamma\right)S_a=-2\alpha\epsilon h^c_a\nabla_c\left[R\right],
\end{equation}
\begin{equation}\label{dls2}
\left(8\pi+\gamma\right)S=2\alpha K^\Sigma\left[R\right],
\end{equation}
\begin{equation}\label{dldsab2}
\xi_{ab}\left(l\right)\equiv\left(8\pi+\gamma\right)s_{ab}\left(l\right)+\frac{1}{2}h_{ab}\gamma s_c^c\left(l\right)=2\alpha\epsilon\Omega^R\left(l\right),
\end{equation}
where the distribution function $\xi_{ab}\left(l\right)$ just defined can also be written in an explicit form for some test function $Y^{ab}$ as
\begin{equation}
\int_\Omega \xi_{ab}Y^{ab}d^4x=-\int_\Sigma 2\alpha\epsilon h_{ab}\left[R\right]n^c\nabla_cY^{ab}d^3x.
\end{equation}
As we have chosen the function $f\left(R,T\right)$ wisely to avoid any further restrictions in the trace of the stress-energy tensor $T_{ab}$, in this case a gravitational double-layer will be present at the hypersurface $\Sigma$ and it is described by the set of Eqs.\eqref{dlsab2} to \eqref{dldsab2}. These equations, along with Eqs.\eqref{junctiondl2.1} and \eqref{junctiondl2.2} constitute the full set of junction conditions of the $f\left(R,T\right)$ gravity in the particular case of a function of the form of Eq.\eqref{functiondl2}, and it is the most general case for which the gravitational double-layers arise. Every extra term arising in this situation is proportional to $\left[R\right]$ as expected. The general results from Sec.\ref{sec:jcts1} can thus be obtained by taking the limit $\left[R\right]=\left[T\right]=0$. A similar set of conditions has also been obtained in $f\left(R\right)$ gravity for the particular case of timelike hypersurfaces and our results can be matched to those in $f\left(R\right)$ by taking $\epsilon=1$ and $\gamma=0$.

\section{Junction conditions in the scalar-tensor representation}\label{sec:juncsca}

\subsection{matching with a thin-shell at $\Sigma$}\label{sec:jcts2}

Let us now turn to the scalar-tensor representation of the theory derives in Sec.\ref{sec:scaten}. The method to derive the first junction condition in this representation is completely analogous to the one followed in the geometrical representation of the theory. We start by writing the metric in the distribution formalism as
\begin{equation}\label{metric2}
g_{ab}=g_{ab}^+\Theta\left(l\right)+g_{ab}^-\Theta\left(-l\right).
\end{equation}
The partial derivatives of Eq.\eqref{metric2} are thus $\partial_cg_{ab}=\partial_c g_{ab}^+\Theta\left(l\right)+\partial_cg_{ab}^-\Theta\left(-l\right)+\epsilon\left[g_{ab}\right]n_c\delta\left(l\right)$. When one defines the Christoffel symbols associated to this metric, one again needs to avoid the presence of products in the form $\Theta\left(l\right)\delta\left(l\right)$, as they are undefined in the distribution formalism. As a consequence, one imposes $\left[g_{ab}\right]=0$. Furthermore, as $g_{ab}$ induces a metric on $\Sigma$ given by $h_{\alpha\beta}=e^a_\alpha e^b_\beta g_{ab}$, to preserve the continuity of the metric at $\Sigma$, the same result must hold for the induced metric $h_{\alpha\beta}$, i.e., the first junction condition takes again the form
\begin{equation}\label{junction3.1}
\left[h_{\alpha\beta}\right]=0.
\end{equation}
This result is consistent with the geometrical representation, as Eq.\eqref{junction3.1} is the same as Eq.\eqref{junction1.1}. Taking Eq.\eqref{junction3.1} into account, the partial derivatives of the metric $g_{ab}$ become
\begin{equation}\label{dmetric2}
\partial_cg_{ab}=\partial_c g_{ab}^+\Theta\left(l\right)+\partial_cg_{ab}^-\Theta\left(-l\right).
\end{equation}
Using Eq.\eqref{dmetric2}, one can now construct the Christoffel symbols for the metric $g_{ab}$ in the distribution formalism without undefined terms, and consequently one is also able to compute the Ricci tensor $R_{ab}$ and the Ricci scalar $R$. These two quantities can be written in the distribution formalism generally as
\begin{eqnarray}\label{rab3.1}
R_{ab}&=&R_{ab}^+\Theta\left(l\right)+R_{ab}^-\Theta\left(-l\right)-\nonumber \\
&-&\left(\epsilon e_a^\alpha e_b^\beta
\left[K_{\alpha\beta}\right]+n_an_b\left[K\right]\right)\delta\left(l\right),
\end{eqnarray}
\begin{equation}\label{ricci3.1}
R=R^+\Theta\left(l\right)+R^-\Theta\left(-l\right)-2\epsilon\left[K\right]\delta\left(l\right).
\end{equation}
where $K_{\alpha\beta}=\nabla_\alpha n_\beta$ is the extrinsic curvature of the hypersurface $\Sigma$ where $n_\beta=e^b_\beta n_b$, and $K=K^\alpha_\alpha$ is the corresponding trace. 

Let us now turn to the matter sector of the theory. In the previous paragraphs, more specifically in Eqs.\eqref{rab3.1} and \eqref{ricci3.1}, we have shown that the Ricci tensor $R_{ab}$ and the Ricci scalar $R$ present terms proportional to $\delta\left(l\right)$, which will appear in the left-hand side of the field equations in Eq.\eqref{fieldst}. These terms can be associated with the presence of a thin-shell of matter at the separation hypersurface $\Sigma$. To find the properties of the thin-shell, let us write the stress-energy tensor $T_{ab}$ as a distribution function of the form
\begin{equation}\label{set3.1}
T_{ab}=T_{ab}^+\Theta\left(l\right)+T_{ab}^-\Theta\left(-l\right)+\delta\left(l\right)S_{ab},
\end{equation} 
where $S_{ab}$ is the four-dimensional stress-energy tensor of the thin-shell, which can be written as a three-dimensional tensor at $\Sigma$ as
\begin{equation}
S_{ab}=S_{\alpha\beta}e^\alpha_ae^\beta_b.
\end{equation}
The field equations in Eq.\eqref{fieldst} and the scalar field equation in Eq. \eqref{eompsi} also depend explicitly in the trace of the stress-energy tensor $T=g^{ab}T_{ab}$. Taking the trace of Eq.\eqref{set3.1}, we find that $T$ becomes
\begin{equation}\label{T3.1}
T=T^+\Theta\left(l\right)+T^-\Theta\left(-l\right)+\delta\left(l\right)S,
\end{equation}
where we have defined $S=S_a^a$. Up to this point there are still no restrictions in the divergent terms of the Ricci tensor and scalar $R_{ab}$ and $R$ or the stress-energy tensor $T_{ab}$ and its trace $T$. 

Consider now the contribution of the scalar fields $\varphi$ and $\psi$. In the distribution formalism, one writes the scalar fields in the usual way as
\begin{equation}\label{phi3.1}
\varphi=\varphi^+\Theta\left(l\right)+\varphi^-\Theta\left(-l\right),
\end{equation}
\begin{equation}\label{psi3.1}
\psi=\psi^+\Theta\left(l\right)+\psi^-\Theta\left(-l\right).
\end{equation}
As the scalar fields $\varphi$ and $\psi$ are defined with no dependency in the $\delta\left(l\right)$ distribution, it is already guaranteed that the potential function $V\left(\varphi,\psi\right)$ will be regular and well-defined, as products of the form $\Theta\left(l\right)\delta\left(l\right)$ or $\delta^2\left(l\right)$ will not arise. As a consequence, the same regularity will be also guaranteed in any of the partial derivatives of $V$. In particular, the left-hand side of Eqs.\eqref{eomphi} and \eqref{eompsi}, which are $V_\varphi$ and $V_\psi$ respectively, will not have any dependency on $\delta\left(l\right)$. Consequently, from Eqs.\eqref{eomphi} and \eqref{eompsi}, one concludes that the Ricci scalar $R$ and the trace of the stress-energy tensor $T$ must not have any dependence on $\delta\left(l\right)$. Considering the explicit forms of these two variables in Eqs.\eqref{ricci3.1} and \eqref{T3.1} respectively, one derives the second and third junction conditions:
\begin{equation}\label{junction3.2}
\left[K\right]=0,
\end{equation}
\begin{equation}\label{junction3.3}
S=0.
\end{equation}
These results are consistent with the ones previously obtained in the geometrical representation of the theory, as they were also derived in Eqs.\eqref{junction1.2} and \eqref{junction1.4}. The Ricci scalar $R$ and the trace of the stress-energy tensor $T$ become thus
\begin{equation}\label{ricci3.2}
R=R^+\Theta\left(l\right)+R^-\Theta\left(-l\right),
\end{equation}
\begin{equation}\label{T3.2}
T=T^+\Theta\left(l\right)+T^-\Theta\left(-l\right),
\end{equation}
and the Ricci tensor $R_{ab}$ simplifies to
\begin{equation}\label{rab3.2}
R_{ab}=R_{ab}^+\Theta\left(l\right)+R_{ab}^-\Theta\left(-l\right)-\epsilon e_a^\alpha e_b^\beta
\left[K_{\alpha\beta}\right]\delta\left(l\right).
\end{equation}

As the field equations in Eq.\eqref{fieldst} depend on second order derivatives of the scalar field $\varphi$, one has to examine these terms in the distribution formalism as well. Taking the partial derivative of Eq.\eqref{phi3.1}, one obtains
\begin{equation}\label{dvarphi3.1}
\partial_a\varphi=\varphi^+\Theta\left(l\right)+\varphi^-\Theta\left(-l\right)+\epsilon\left[\varphi\right]n_a\delta\left(l\right).
\end{equation}
In a general scalar-tensor theory of gravity with a scalar field $\phi$, e.g., Brans-Dicke theory with a parameter $\omega_{BD}\neq 0$, it is the presence of a kinetic term that forces the scalar fields to be continuous, due to the presence of undefined terms of the form $\Theta\left(l\right)\delta\left(l\right)$ and divergent terms $\delta^2\left(l\right)$ in the products $\partial_a\phi\partial_b\phi$. However, in the absence of a kinetic term, there are no reasons \textit{a priori} for the condition $\left[\varphi\right]$ to be mandatory. However, in this work we are not considering a general scalar-tensor theory represented by the action in Eq.\eqref{actionst}, but rather with an equivalent scalar-tensor representation of the $f\left(R,T\right)$ theory described by the action in Eq.\eqref{action}. As explained in Sec. \ref{sec:scaten}, this representation is only defined whenever the determinant of the matrix $\mathcal M$ in Eq.\eqref{matrixeq} does not vanish. This property implies that it must be possible to write the scalar fields $\varphi$ and $\psi$ explicitly in terms of $R$ anf $T$, i.e., $\varphi=\varphi\left(R,T\right)$ and $\psi=\psi\left(R,T\right)$, and vice-versa, i.e., $R=R\left(\varphi,\psi\right)$ and $T=T\left(\varphi,\psi\right)$. Taking these arguments into consideration, one can write the first and second-order covariant derivatives of the scalar field $\varphi$ as
\begin{equation}\label{dvarphiRT}
\partial_a \varphi= \varphi_{R}\partial_aR+\varphi_{T}\partial_aT,
\end{equation}
\begin{eqnarray}
&&\nabla_a\nabla_b\varphi=\varphi_{R}\nabla_a\nabla_bR+\varphi_{T}\nabla_a\nabla_bT+\label{ddvarphiRT}\\
&&+\varphi_{RR}\partial_aR\partial_bR+\varphi_{TT}\partial_aT\partial_bT+2\varphi_{RT}\partial_{(a}R\partial_{b)}T,\nonumber
\end{eqnarray}
where the subscripts $R$ and $T$ denote partial derivatives with respect to these quantities, respectively. Taking the partial derivatives of $R$ and $T$ from Eqs.\eqref{ricci3.2} and \eqref{T3.2} respectively, one obtains in the distribution formalism
\begin{equation}\label{dricci3.1}
\partial_a R=\partial_aR^+\Theta\left(l\right)+\partial_aR^-\Theta\left(-l\right)+\epsilon\left[R\right]n_a\delta\left(l\right),
\end{equation}
\begin{equation}\label{dt3.1}
\partial_a T=\partial_aT^+\Theta\left(l\right)+\partial_aT^-\Theta\left(-l\right)+\epsilon\left[T\right]n_a\delta\left(l\right).
\end{equation}
Thus, the presence of products of the form $\partial_aR\partial_bR$, $\partial_aT\partial_bT$ and $\partial_aR\partial_bT$ in the expression for $\nabla_a\nabla_b\varphi$ in Eq.\eqref{ddvarphiRT} implies that these differential terms will depend on products of the form $\Theta\left(l\right)\delta\left(l\right)$ and $\delta\left(l\right)^2$, which are undefined and singular, respectively. To avoid the presence of these terms, one must force the $\delta\left(l\right)$ terms in Eqs.\eqref{dricci3.1} and \eqref{dt3.1} to vanish, i.e., $\left[R\right]=0$ and $\left[T\right]=0$. Since both the scalar fields are well-behaved functions of $R$ and $T$ by the definition of the equivalent scalar-tensor representation as explained earlier in this section, the conditions $\left[R\right]=0$ and $\left[T\right]=0$ imply the fourth and fifth junction conditions:
\begin{equation}\label{junction3.4}
\left[\varphi\right]=0,
\end{equation}
\begin{equation}\label{junction3.5}
\left[\psi\right]=0.
\end{equation}
Junction conditions of this form are common in general scalar-tensor theories of gravity as long as the action features a kinetic term for the scalar field. However, here they arise even though the action in Eq.\eqref{actionst} does not have a kinetic term for either $\varphi$ and $\psi$, due to the very definition of the equivalent scalar-tensor representation. A similar situation arises in the metric formalism of $f\left(R\right)$ gravity and the hybrid metric-Palatini gravity. As a consequence, the first-order derivative of $\varphi$ in Eq.\eqref{dvarphi3.1} becomes
\begin{equation}\label{dvarphi3.2}
\partial_a\varphi=\varphi^+\Theta\left(l\right)+\varphi^-\Theta\left(-l\right),
\end{equation}
and we are finally able to compute the second-order derivatives of the scalar field $\varphi$ which become generally
\begin{eqnarray}
\nabla_a\nabla_b\varphi&=&\nabla_a\nabla_b\varphi^+\Theta\left(l\right)+\nabla_a\nabla_b\varphi^-\Theta\left(-l\right)+\nonumber \\
&+&\epsilon n_a\left[\partial_b\varphi\right]\delta\left(l\right).\label{ddvarphi3.1}
\end{eqnarray}

We are now in conditions of deriving the stress-energy tensor of the thin shell $S_{\alpha\beta}$. To do so, one introduces the representations of the various quantities in the distribution formalism given by Eqs.\eqref{metric2}, \eqref{set3.1}, \eqref{phi3.1}, \eqref{psi3.1}, \eqref{ricci3.2}, \eqref{T3.2}, \eqref{rab3.2}, and \eqref{ddvarphi3.1} into the field equations in Eq.\eqref{fieldstpf} and project the result into the hypersurface $\Sigma$ with $e^a_\alpha e^b_\beta$. Defining $\varphi|_\Sigma=\varphi_\Sigma$ and $\psi|_\Sigma=\psi_\Sigma$ as the values of the scalar fields at $\Sigma$, one obtains
\begin{equation}\label{sab3.1}
\left(8\pi+\psi_\Sigma\right)S_{\alpha\beta}=-\epsilon \varphi_\Sigma\left[K_{\alpha\beta}\right]+\epsilon h_{\alpha\beta}n^c\left[\partial_c\varphi\right].
\end{equation}
Taking the trace of Eq.\eqref{sab3.1}, inserting the result into Eq.\eqref{junction3.3} and simplifying the outcome with Eq.\eqref{junction3.2}, yields a more convenient form of the third junction condition as $\left[\partial_a\varphi\right]=0$, which can then be reinserted into Eq.\eqref{sab3.1} to force the second term on the right-hand side to vanish.

To summarize, the complete set of junction conditions for the equivalent scalar-tensor representation of the $f\left(R,T\right)$ gravity in for the general matching in the presence of a thin-shell at $\Sigma$ is composed of a total of six equations of the form
\begin{eqnarray}
&\left[h_{\alpha\beta}\right]=0, \nonumber \\
&\left[K\right]=0, \nonumber \\
&\left[\varphi\right]=0, \label{fullset3} \\
&\left[\psi\right]=0, \nonumber \\
&\left[\partial_a\varphi\right]=0, \nonumber \\
&\left(8\pi + \psi_\Sigma\right)S_{\alpha\beta}=-\epsilon\varphi_\Sigma\left[K_{\alpha\beta}\right].\nonumber
\end{eqnarray}
Note that this set of junction conditions could be derived directly from the system in Eq.\eqref{fullset1} via the introduction of the definitions $\varphi=f_R\left(R,T\right)$ and $\psi=f_T\left(R,T\right)$, which emphasizes the equivalence between the two representations of the theory.

\subsection{Smooth matching at $\Sigma$}

In the previous section we have extended the analysis of Sec. \ref{sec:jcts1} to the equivalent scalar-tensor representation of the $f\left(R,T\right)$ theory. In particular, we have considered the matching of two spacetimes $\mathcal V^\pm$ at a separation hypersurface $\Sigma$ allowing for a thin-shell of matter described by a stress-energy tensor $S_{\alpha\beta}$ to exist at $\Sigma$. Similarly as before, if one is interested in a smooth matching instead, one has to pursue the same analysis but for a vanishing $S_{\alpha\beta}$, for which a new set of junction conditions will arise. The vanishing of $S_{\alpha\beta}$ is guaranteed by forcing all the terms proportional to $\delta\left(l\right)$ that could possibly appear in the field equations of Eq.\eqref{fieldstpf} to vanish.

Let us start from the metric $g_{ab}$ again. In this situation, the metric provided in Eq.\eqref{metric2} in the distribution formalism does not vary when one considers a smooth matching, once it does not have any dependence in $\delta\left(l\right)$. Consequently, one can follow the same reasoning as in Sec.\ref{sec:jcts2} to conclude that the induced metric $h_{\alpha\beta}$ must remain continuous across $\Sigma$, and the first junction condition becomes
\begin{equation}\label{junction4.1}
\left[h_{\alpha\beta}\right]=0.
\end{equation}
One can now proceed to the calculation of the Christoffel symbols, the Ricci tensor $R_{ab}$ and the Ricci scalar $R$ associated to this metric, which are given by the general forms provided in Eqs.\eqref{rab3.1} and \eqref{ricci3.1}, respectively, where $K_{\alpha\beta}=\nabla_\alpha n_\beta$ represents the extrinsic curvature of the hypersurface $\Sigma$. Given the presence of a term proportional to $R_{ab}$ in the field equations in Eq.\eqref{fieldstpf}, the term proportional to $\delta\left(l\right)$ in Eq.\eqref{rab3.1} must be forced to vanish for the matching to be smooth, from which one concludes that $K_{\alpha\beta}$ must be continuous across $\Sigma$, i.e., the second junction condition becomes
\begin{equation}\label{junction4.2}
\left[K_{\alpha\beta}\right],
\end{equation}
which is consistent with what was already obtained in the geometrical representation of the theory in Eq.\eqref{junction2.2}. As Eq.\eqref{junction4.2} features $\left[K\right]=0$ as a consequence, the latter does not have to be imposed separately. The Ricci tensor $R_{ab}$ and the Ricci scalar $R$ thus become
\begin{equation}\label{rab4.1}
R_{ab}=R_{ab}^+\Theta\left(l\right)+R_{ab}^-\Theta\left(-l\right),
\end{equation}
\begin{equation}\label{ricci4.1}
R=R^+\Theta\left(l\right)+R^-\Theta\left(-l\right).
\end{equation}

Turning now to the matter sector of the theory, and since we are interested in matching the two spacetimes smoothly at $\Sigma$, the stress-energy tensor $T_{ab}$ but not have any dependence on $\delta\left(l\right)$. Similarly as before, this situation forces $T_{ab}$ and its trace $T$ to take the forms
\begin{equation}\label{set4}
T_{ab}=T_{ab}^+\Theta\left(l\right)+T_{ab}^-\Theta\left(-l\right),
\end{equation}
\begin{equation}\label{T4}
T=T^+\Theta\left(l\right)+T^-\Theta\left(-l\right).
\end{equation}

Finally, let us now analyze the contribution of the scalar fields. Again, one writes the scalar fields $\varphi$ and $\psi$ in the distribution formalism in the same forms as provided in Eqs.\eqref{phi3.1} and \eqref{psi3.1}, respectively. These scalar fields are defined with no dependencies in $\delta\left(l\right)$, and thus the potential $V\left(\varphi,\psi\right)$ is already guaranteed to be well-defined and regular. This regularity is preserved upon partial derivatives of $V$ with respect to the scalar fields $\varphi$ and $\psi$. Consequently, the equations of motion for $\varphi$ and $\psi$, i.e., Eqs.\eqref{eomphi} and \eqref{eompsi}, whose left-hand sides depend on $V_\varphi$ and $V_\psi$ respectively, are automatically satisfied at the hypersurface $\Sigma$, as the regularity of $R$ and $T$ is forced by Eqs.\eqref{ricci4.1} and \eqref{T4}.

To analyze the contribution of the differential terms $\nabla_a\nabla_b\varphi$ and $\Box\varphi$ in the field equations in Eq.\eqref{fieldstpf}, one takes the derivatives of the scalar field $\varphi$, which will be of the same form as provided in Eq.\eqref{dvarphi3.1}. Following the same arguments of Sec.\eqref{sec:jcts2}, i.e., the fact that the scalar fields are not arbitrary functions but must be constrained by the condition for which the scalar-tensor representation of the theory is defined (see Sec.\ref{sec:scaten} for more details), and the fact that, as a consequence of this condition, the scalar fields $\varphi$ and $\psi$ must be well-behaved functions of $R$ and $T$, one concludes that to avoid the presence of undefined products $\Theta\left(l\right)\delta\left(l\right)$ or singular products $\delta\left(l\right)^2$ in the second order derivatives of $\varphi$ given in Eq.\eqref{ddvarphiRT} one must impose that $\left[R\right]=0$ and $\left[T\right]=0$. Therefore, following that $\varphi=\varphi\left(R,T\right)$ and $\psi=\psi\left(R,T\right)$, one recovers the third and fourth junction conditions as
\begin{equation}\label{junction4.3}
\left[\varphi\right]=0,
\end{equation}
\begin{equation}\label{junction4.4}
\left[\psi\right]=0.
\end{equation}
Again, we emphasize that junction conditions of this form are common in scalar-tensor theories featuring kinetic terms for the scalar fields, but in the case of the scalar-tensor representation of $f\left(R,T\right)$ these conditions are imposed by the constraints on the scalar fields needed for the scalar-tensor representation to be well-defined.

Let us now turn to the second-order differential terms $\nabla_a\nabla_bR$ and $\nabla_a\nabla_bT$ in Eq.\eqref{ddvarphiRT}. Taking the second-order covariant derivatives of $R$ and $T$ in Eqs.\eqref{ricci4.1} and \eqref{T4}, and taking into consideration that $\left[R\right]=0$ and $\left[T\right]=0$, one obtains the same results as in Eqs.\eqref{ddricci1.1} and \eqref{ddT1.1}, respectively. These forms feature terms proportional to $\delta\left(l\right)$, which consequently will be present in $\nabla_a\nabla_b\varphi$ via Eq.\eqref{ddvarphiRT}. In the previous section, where we have considered a matching with a thin-shell at $\Sigma$, these terms were not problematic as they could be absorbed into the stress-energy tensor $S_{ab}$ of the thin-shell. In this section however, we are interested in guaranteeing a smooth matching between the two spacetimes, and thus these terms proportional to $\delta\left(l\right)$ must not be present in the field equations in Eq.\eqref{fieldstpf}. Thus, we have to force $\left[\partial_aR\right]=0$ and $\left[\partial_aT\right]=0$. Since the scalar fields $\varphi$ and $\psi$ must be well-behaved functions of $R$ and $T$ for the scalar-tensor representation of the theory to be well-defined, their first-order partial derivatives are
\begin{equation}
\partial_a\varphi=\varphi_R\partial_aR+\varphi_T\partial_aT,
\end{equation}
\begin{equation}
\partial_a\psi=\psi_R\partial_aR+\psi_T\partial_aT,
\end{equation}
from which it becomes clear that, if the first-order derivatives of $R$ and $T$ are continuous, so must be the first-order derivatives of $\varphi$ and $\psi$. These considerations thus translate into the fifth and sixth junction conditions of the form
\begin{equation}\label{junction4.5}
\left[\partial_a\varphi\right]=0,
\end{equation}
\begin{equation}\label{junction4.6}
\left[\partial_a\psi\right]=0.
\end{equation}

To summarize, the complete set of junction conditions for the equivalent scalar-tensor representation of the $f\left(R,T\right)$ gravity in the particular case of a smooth matching between two spacetimes at $\Sigma$ is thus composed of the following six equations
\begin{eqnarray}
&\left[h_{\alpha\beta}\right]=0, \nonumber \\
&\left[K_{\alpha\beta}\right]=0, \nonumber \\
&\left[\varphi\right]=0, \label{fullset4} \\
&\left[\psi\right]=0, \nonumber \\
&\left[\partial_a\varphi\right]=0, \nonumber \\
&\left[\partial_a\psi\right]=0.
\end{eqnarray}
Note that this set of junction conditions could be derived directly from the system in Eq.\eqref{fullset2} via the introduction of the definitions $\varphi=f_R\left(R,T\right)$ and $\psi=f_T\left(R,T\right)$, which emphasizes the equivalence between the two representations of the theory.

\subsection{Double gravitational layers at $\Sigma$}

\subsubsection{Matching with $\left[\psi\right]\neq0$}\label{sec:scadl1}

In Sec.\ref{sec:jcts2} we derived the general set of junction conditions necessary to match two spacetimes $\mathcal V^\pm$ at a separation hypersurface $\Sigma$ in the presence of a thin-shell. This analysis was done for a general form of the potential $V\left(\varphi,\psi\right)$, which is associated to a general function $f\left(R,T\right)$ via Eq.\eqref{defpot}. In particular, the junction conditions in Eqs.\eqref{junction4.5} and \eqref{junction4.6} were derived as a consequence of the dependence of $\varphi$ and $\psi$ in $R$ and $T$ and the fact that $\left[R\right]=0$ and $\left[T\right]=0$. However, in Sec.\eqref{sec:geodl1}, we have shown that there are particular forms of the function $f\left(R,T\right)$ for which the latter conditions can be discarded. Analogously, one expects that for some particular forms of the potential $V\left(\varphi,\psi\right)$ the junction conditions $\left[\varphi\right]=0$ and $\left[\psi\right]=0$ can be discarded. 

Since the scalar-tensor representation derived in Sec.\eqref{sec:scaten} is only well-defined when the determinant of the matrix $\mathcal M$ defined in Eq.\eqref{matrixeq} is non-vanishing, this condition will impose a constraint in the function $f\left(R,T\right)$ to be used in this section. Taking the second-order partial derivatives with respect to $R$ and $T$ of the function provided in Eq.\eqref{functiondl} and computing the determinant of $\mathcal M$, one verifies that the parameters $\alpha$, $\gamma$ and the function $g\left(T\right)$ are constrained by the condition
\begin{equation}\label{conddl}
2\alpha g''\left(T\right)-\gamma^2\neq 0.
\end{equation}

The form of the potential $V\left(\varphi,\psi\right)$ associated to the function $f\left(R,T\right)$ in Eq.\eqref{functiondl} can be obtained by computing the partial derivatives of $f\left(R,T\right)$, using the definitions of the scalar fields $\varphi$ and $\psi$ in Eq.\eqref{defsca} to invert these relations and obtain $R\left(\varphi,\psi\right)$ and $T\left(\varphi,\psi\right)$, and finally introducing the results into Eq.\eqref{defpot}. We thus obtain
\begin{eqnarray}\label{potdl}
V\left(\varphi,\psi\right)&=&-R\left(\varphi,\psi\right)\left[1+\gamma T\left(\varphi,\psi\right)\right]+2\Lambda-\alpha R\left(\varphi,\psi\right)^2-\nonumber \\
&-&g\left[T\left(\varphi,\psi\right)\right]+\varphi R\left(\varphi,\psi\right)+\psi T\left(\varphi,\psi\right),
\end{eqnarray}
\begin{equation}\label{Rinv}
R\left(\varphi,\psi\right)=\frac{1}{2\alpha}\left(\varphi-1\right)-\frac{\gamma}{2\alpha}h^{-1}\left[\psi-\frac{\gamma}{2\alpha}\left(\varphi-1\right)\right],
\end{equation}
\begin{equation}\label{Tinv}
T\left(\varphi,\psi\right)=h^{-1}\left[\psi-\frac{\gamma}{2\alpha}\left(\varphi-1\right)\right],
\end{equation}
where $h^{-1}$ is the inverse of a function $h\left(T\right)$ defined as
\begin{equation}\label{deffunh}
h\left(T\right)=g'\left(T\right)-\frac{\gamma^2}{2\alpha}T.
\end{equation}
Note that, given the constraint provided in Eq.\eqref{conddl}, the function $h\left(T\right)$ is also constrained to satisfy $h'\left(T\right)\neq 0$, which is a necessary condition to guarantee its invertibility. These results are true for any well-behaved function $g\left(T\right)$. Choosing an explicit form of the function $g\left(T\right)$ will thus set an explicit form for the function $h\left(T\right)$ via Eq.\eqref{deffunh}, which consequently allows us to find $R$ and $T$ from Eqs.\eqref{Rinv} and \eqref{Tinv}, respectively, and finally an explicit form of the potential from Eq.\eqref{potdl}.

Let us now turn to the junction conditions arising from a theory with a potential described by Eq.\eqref{potdl}. The metric $g_{ab}$ and its partial derivatives are still given in the form of Eqs.\eqref{metric2} and \eqref{dmetric2} respectively, and the Ricci scalar and its partial derivatives are given by Eqs.\eqref{ricci3.1} and \eqref{dricci3.1} respectively. The analysis that leads to Eqs.\eqref{junction3.1} and \eqref{junction3.2} can be followed integrally and thus the first and second junction conditions remain as
\begin{equation}\label{junctiondl3.1}
\left[h_{\alpha\beta}\right]=0,
\end{equation}
\begin{equation}\label{junctiondl3.2}
\left[K\right]=0.
\end{equation}

In the matter sector, since the potential $V\left(\varphi,\psi\right)$ given in Eq.\eqref{potdl} depends on an arbitrary function $g\left[T\left(\varphi,\psi\right)\right]$, which can depend generally power-laws of $T$, and also in products of $R\left(\varphi,\psi\right)T\left(\varphi,\psi\right)$, one concludes that $T$ can not have any dependence on $\delta\left(l\right)$, otherwise undefined products $\Theta\left(l\right)\delta\left(l\right)$ or singular products $\delta\left(l\right)^2$ would appear in the potential and, consequently, in the field equations in Eq.\eqref{fieldst}. Thus, the trace of the stress-energy tensor $T$ must be written in the form of Eq.\eqref{T3.2}, and whichever terms proportional to $\delta\left(l\right)$ that arise in $T_{ab}$ must vanish upon tracing. Also, since the potential $V$ depends in general of products and power-laws of $\varphi$ and $\psi$, one must guarantee that $\varphi$ and $\psi$ are still given by Eqs.\eqref{phi3.1} and \eqref{psi3.1}, respectively, as to avoid the same problematic products in the field equations. These considerations guarantee the regularity of the potential $V$ and its partial derivatives with respect to $\varphi$ and $\psi$, and the equations of motion for the scalar fields deduced in Eqs.\eqref{eomphi} and \eqref{eompsi} are automatically well-behaved at $\Sigma$.

The procedure to find the remaining junction conditions in this case are the same up until the contribution of the second-order derivatives of $\varphi$, i.e., the terms $\nabla_a\nabla_b\varphi$ and $\Box\varphi$. In this case the scalar fields $\varphi$ and $\psi$ can be written in terms of $R$ and $T$ as $\varphi=f_R=1+\gamma T+2\alpha R$ and $\psi=\gamma R+g'\left(T\right)$, which implies that the second order derivatives of $\varphi$ given in Eq.\eqref{ddvarphiRT} become
\begin{equation}
\nabla_a\nabla_b\varphi=2\alpha\nabla_a\nabla_bR+\gamma\nabla_a\nabla_bT.
\end{equation}
In the general case of Sec.\eqref{sec:jcts2}, the continuity of $R$ and $T$ was mandatory due to the existence of products between $\partial_aR$ and $\partial_aT$ in $\nabla_a\nabla_v\varphi$, which are now absent since $\varphi_{RR}=\varphi_{TT}=\varphi_{RT}=0$. Thus, the particular case in study allows for $\left[R\right]\neq 0$ and $\left[T\right]\neq=0$. Since the scalar fields $\varphi$ and $\psi$ are written in terms of $R$ and $T$, this result implies that the junction conditions $\left[\varphi\right]=0$ and $\left[\psi\right]=0$ from the full set of Eq.\eqref{fullset3} can be discarded. The first and second order covariant derivatives of the scalar field $\varphi$ thus become
\begin{equation}\label{dphidl3.1}
\partial_a\varphi=\partial_a\varphi^+\Theta\left(l\right)+\partial_a\varphi^-\Theta\left(-l\right)+\epsilon n_a\left[\varphi\right]\delta\left(l\right),
\end{equation}
\begin{equation}\label{ddphidl3.1}
\nabla_a\nabla_b\varphi=\left(\nabla^2\varphi\right)_{ab}+\epsilon\nabla_a\left(\left[\varphi\right]\delta\left(l\right)n_b\right),
\end{equation}
where $\left(\nabla^2\varphi\right)_{ab}$ collectively denotes the right-hand side of Eq.\eqref{ddvarphi3.1}. Similarly to what happens in the geometrical representation of the theory, the scalar-tensor representation allows for one to discard a couple of junction conditions, in this case $\left[\varphi\right]\neq 0$ and $\left[\psi\right]\neq0$ are allowed at this point, but at the cost of extra terms in the stress-energy tensor $S_{\alpha\beta}$ of the thin-shell. The second term in the right-hand side of Eq.\eqref{ddphidl3.1} can be written explicitly as
\begin{eqnarray}
&&\nabla_a\left(\left[\varphi\right]\delta\left(l\right)n_b\right)=\Delta_{ab}^\varphi+\nonumber \\
&&+\delta\left(l\right)\left(K_{ab}-\epsilon K n_an_b+n_bh_a^c\nabla_c\right)\left[\varphi\right],\label{dlddphi}
\end{eqnarray}
where the distribution function $\Delta_{ab}^\varphi$ is defined as
\begin{equation}
\int_\Omega\Delta_{ab}^\varphi Y^{ab}d^4x=-\epsilon\int_\Sigma\left[\varphi\right]n_a n_bn^c\nabla_cY^{ab}d^3x,
\end{equation}
for a given test function $Y^{ab}$. Inserting Eqs.\eqref{dlddphi} and Eq.\eqref{ddvarphi3.1} into Eq.\eqref{ddphidl3.1} and inserting the result back into the field equations in Eq.\eqref{fieldst}, one verifies that the stress-energy tensor $T_{ab}$ can be recast into the form
\begin{eqnarray}
T_{ab}&=&T_{ab}^+\Theta\left(l\right)+T_{ab}^-\Theta\left(-l\right)+\label{dlTab2}\\
&+&\delta\left(l\right)\left(S_{ab}+2S_{(a}n_{b)}+Sn_an_b\right)+s_{ab}\left(l\right),\nonumber
\end{eqnarray}
where the stress-energy tensor of the thin-shell $S_{ab}$, the external momentum flux $S_a$, the external normal pressure $S$, and the double-layer stress-energy tensor distribution $s_{ab}$ are written in terms of the geometrical quantities in the forms
\begin{equation}\label{dlstSab}
\left(8\pi+\psi_\Sigma\right)S_{ab}=-\epsilon\varphi_\Sigma\left[K_{ab}\right]+\epsilon h_{ab}n^c\left[\nabla_c\varphi\right]-\epsilon K_{ab}^\Sigma\left[\varphi\right]
\end{equation}
\begin{equation}\label{dlstSa}
\left(8\pi+\psi_\Sigma\right)S_a=-\epsilon h_a^c\nabla_c\left[\varphi\right]
\end{equation}
\begin{equation}\label{dlstS}
\left(8\pi+\psi_\Sigma\right)S=K^\Sigma\left[\varphi\right]
\end{equation}
\begin{equation}\label{dlstsab}
\left(8\pi+\psi_\Sigma\right)s_{ab}\left(l\right)=h_{ab}\Delta^\varphi-\Delta_{ab}^\varphi,
\end{equation}
where we have defined $\varphi_\Sigma$ and $\psi_\Sigma$ as the average values of the scalar fields at the hypersurface $\Sigma$, i.e., $2\varphi_\Sigma=\varphi^++\varphi^-$ and $2\psi_\Sigma=\psi^++\psi^-$, and $\Delta^\varphi$ is the trace od $\Delta_{ab}^\varphi$. One can thus express the double-layer stress-energy tensor distribution explicitly as
\begin{equation}\label{dlstintsab}
\int_\Omega \left(8\pi + \psi_\Sigma\right) s_{ab}Y^{ab}d^4x=-\int_\Sigma \epsilon h_{ab}\left[\varphi\right]n^c\nabla_cY^{ab}d^3x.
\end{equation}

Previously, to guarantee the regularity of the potential $V\left(\varphi\psi\right)$, we have forced the trace of the stress-energy tensor $T$ to be written as in Eq.\eqref{T3.2}. This condition will impose constraints on Eq.\eqref{dlTab2}. Taking the trace of Eq.\eqref{dlTab2}, simplifying the result using Eqs.\eqref{junctiondl3.2}, \eqref{dlstSa}, \eqref{dlstS} and \eqref{dlstsab}, and using the result $n^ah_a^b=0$, we obtain
\begin{equation}\label{dlstT}
T=T_{ab}^+\Theta\left(l\right)+T_{ab}^-\Theta\left(-l\right)+3\epsilon\delta\left(l\right)n^c\left[\nabla_c\varphi\right]+s_a^a\left(l\right).
\end{equation}
We can now compare Eq.\eqref{dlstT} with Eq.\eqref{T3.2} to conclude that both the terms proportional to $\delta\left(l\right)$ and proportional to $s_{ab}\left(l\right)$ must vanish identically to guarantee the regularity of the potential $V$. From these considerations, the third and fourth junction conditions arise as
\begin{equation}\label{junctiondl3.3}
\left[\nabla_c\varphi\right]=0,
\end{equation}
\begin{equation}\label{junctiondl3.4}
\left[\varphi\right]=0.
\end{equation}
These two junction conditions also arise in the general case of an arbitrary potential $V$ or, in other words, for an arbitrary function $f\left(R,T\right)$, as can be seen in the full system of junction conditions for that case in Eq.\eqref{fullset3}. Inserting Eqs.\eqref{junctiondl3.3} and \eqref{junctiondl3.4} into Eqs.\eqref{dlstSab}, \eqref{dlstSa}, \eqref{dlstS} and \eqref{dlstintsab}, one verifies that the quantities $S_a$, $S$ and $s_{ab}\left(l\right)$ vanish and $S_{ab}$ reduces to a single term proportional to the jump of the extrinsic curvature.

Summarizing, the full set of junction conditions for the scalar-tensor representation of the $f\left(R,T\right)$ gravity for the particular case where the potential is written in the form of Eq.\eqref{potdl} is composed of the following five equations
\begin{eqnarray}
&\left[h_{\alpha\beta}\right]=0,\nonumber\\
&\left[K\right]=0,\nonumber\\
&\left[\varphi\right]=0,\label{fullsetdl1}\\
&\left[\nabla_c\varphi\right]=0,\nonumber\\
&\left[8\pi+\psi_\Sigma\right]S_{ab}=-\epsilon\varphi_\Sigma\left[K_{ab}\right].\nonumber
\end{eqnarray}
It is thus possible to consider particular forms of the potential $V\left(\varphi,\psi\right)$ for which the junction condition $\left[\psi\right]=0$ can be discarded from the final set of equations. However, as the potential depends on general functions of $T$, the preservation of regularity of these terms does not allow for a gravitational double-layer to arise at $\Sigma$. Instead, the junction conditions $\left[\varphi\right]=0$ and $\left[\nabla_c\varphi\right]=0$ are recovered.

\subsubsection{Matching with a double-layer at $\Sigma$}\label{sec:scadl2}

In this section we will show that, although there are particular forms of the function $f\left(R,T\right)$ for which two spacetimes $\mathcal V^\pm$ can be matched at a separation hypersurface $\Sigma$ with a gravitational double-layer in the geometrical representation of the theory, see Sec.\ref{sec:geodl2}, the same analysis can not be reproduced in the scalar-tensor representation of the theory, as the equivalence between the two representations is not well-defined for these forms of the function.

The scalar-tensor representation of the theory was derived in Sec.\ref{sec:scaten} and was proven to be well-defined only when the determinant of the matrix $\mathcal M$ defined in Eq.\eqref{matrixeq} is non-vanishing. In Sec.\ref{sec:geodl2} we verified that the gravitational double-layers arise only when the function $f\left(R,T\right)$ is given in the form of Eq.\eqref{functiondl2}. For this form of the function, the partial derivatives with respect to $R$ and $T$ become $f_R=1+2\alpha R$ and $f_T=\gamma$, and the second-order derivatives are $f_{RR}=2\alpha$, $f_{TT}=f_{RT}=0$. Consequently, the determinant of the metric $\mathcal M$ is $f_{RR}f_{TT}-f_{RT}=0$. This implies that the relationship between $R$ and $T$ with $\varphi$ and $\psi$ is not unique, and thus not invertible. One concludes that the analysis of the gravitational double-layers can not be pursued in the scalar-tensor representation of the theory, as it is not well-defined.

\section{Examples and applications}\label{sec:apps}

\subsection{Energy conditions of a spherically symmetric perfect-fluid thin-shell}\label{sec:ex1}

In this section we shall briefly analyze the validity of the energy conditions for the matter thin-shell at the separation hypersurface $\Sigma$ in the particular case where the thin-shell is spherically symmetric and well described by a perfect-fluid stress-energy tensor, i.e., we can write the mixed indexes stress.energy tensor $S_\alpha^\beta$ of the shell in the diagonal form
\begin{equation}\label{diagsab}
S_\alpha^\beta=\text{diag}\left(-\sigma,p_t,p_t\right),
\end{equation}
where $\sigma$ is the surface energy density and $p_t$ is the transverse pressure of the thin-shell. Under these assumptions, the energy conditions, more precisely the Null Energy Condition (NEC), the Weak Energy Condition (WEC), the Strong Energy Condition (SEC), and the Dominant Energy Condition (DEC), are given by the following inequalities
\begin{eqnarray}
&\sigma+p_t>0,\label{nec}\\
&\sigma+p_t>0,\quad \sigma>0\label{wec}\\
&\sigma+2p_t>0,\quad \sigma>0\label{sec}\\
&\sigma>|p_t|,\label{dec}
\end{eqnarray}
respectively. In the usual spherical coordinate system $\left(t,r,\theta,\phi\right)$, Eq.\eqref{junction1.2} becomes $\left[K_t^t\right]+\left[K_\theta^\theta\right]+\left[K_\phi^\phi\right]=0$. Spherical symmetry implies that $K_\theta^\theta=K_\phi^\phi$, and thus $\left[K_t^t\right]=-2\left[K_\theta^\theta\right]$. Inserting these considerations and Eq.\eqref{diagsab} into the last of Eq.\eqref{fullset1}, one obtains a relationship between $\sigma$ and $p_t$ in the geometrical representation of $f\left(R,T\right)$ of the form
\begin{equation}\label{sigmapt}
\sigma=2p_t=\frac{\epsilon f_R}{8\pi+f_T}\left[K_t^t\right].
\end{equation}
Following the same reasoning, the equivalent of Eq.\eqref{sigmapt} for the scalar-tensor representation of the theory can be obtained via the insertion of the previous results for $K_{ab}$ and Eq.\eqref{diagsab} into the last of Eq.\eqref{fullset3}, yielding
\begin{equation}\label{sigmaptsca}
\sigma=2p_t=\frac{\epsilon \varphi_\Sigma}{8\pi+\psi_\Sigma}\left[K_t^t\right].
\end{equation}
The results of Eqs.\eqref{sigmapt} and \eqref{sigmaptsca} allow us to verify the validity of the energy conditions. Since $\sigma+p_t=3\sigma/2$ and $|\sigma|>|p_t|$, Eqs.\eqref{nec} to \eqref{dec} will be automatically satisfied whenever $\sigma>0$ and violated otherwise. 

\subsection{Martinez thin-shell: matching an interior Minkowski to an exterior Schwarzschild}\label{sec:ex2}

Let us now consider the matching with a thin-shell between an interior Minkowski spacetime with an exterior Schwarzschild spacetime at a given separation hypersurface $\Sigma$, which stands at a radius $r_\Sigma$ . This situation was considered for the first time by Martinez in Ref. \cite{Martinez:1996ni}, which was one of the pioneer works in thin-shell thermodynamics. The interior and exterior spacetimes are described my the line elements:
\begin{equation}\label{mink1}
ds^2=-dt^2+dr^2+r^2d\Omega^2,
\end{equation}
\begin{equation}\label{schw1}
ds^2=-\left(1-\frac{2M}{r}\right)\alpha dt^2+\left(1-\frac{2M}{r}\right)^{-1}dr^2+r^2d\Omega^2,
\end{equation}
respectively, where $\left(t,r,\theta,\phi\right)$ are the usual spherical coordinates, $\alpha$ is a dimensionless constant that we introduce for later convenience, $M$ is the mass of the Schwarzschild solution, and $d\Omega^2=d\theta^2+\sin^2\theta d\phi^2$ is the solid angle line element. Both spacetimes in Eqs. \eqref{mink1} and \eqref{schw1} are vacuum spacetimes, i.e., the stress-energy tensor $T_{ab}=0$ vanishes, and they are also described by vanishing Ricci tensors $R_{ab}=0$ and, consequently, vanishing Ricci scalars $R=0$.

The first two junction conditions for a matching with a thin-shell in the geometrical and the scalar-tensor representation of the theory are the same, i.e., $\left[h_{\alpha\beta}\right]=0$ and $\left[K\right]=0$ (see the systems of Eqs.\eqref{fullset1} and \eqref{fullset3}). The first junction condition sets a value for the constant $\alpha$ that guarantees that the time coordinate of the two spacetimes in Eqs.\eqref{mink1} and \eqref{schw1} are continuous,
\begin{equation}\label{junctionex1.1}
\alpha=\left(1-\frac{2M}{r_\Sigma}\right)^{-1}.
\end{equation}

Since the two regions $\mathcal V^\pm$ are spherically symmetric, the angular components of the extrinsic curvature are the same, i.e, $K_{\theta\theta}=K_{\phi\phi}$. Thus, the trace of the extrinsic curvature for both regions can be written in the general form $K=K_t^t+2K_\theta^\theta$. The second junction condition thus becomes
\begin{equation}\label{junctionex1.2}
\left[K\right]=-\frac{2}{r}-\frac{3M-2r}{r^2\sqrt{1-\frac{2M}{r}}}=0.
\end{equation}
This condition is a constraint on the radius $r$ at which the matching can be performed. Solving Eq.\eqref{junctionex1.2} for $r$ one finds that the matching is only possible if the hypersurface $\Sigma$ has a radius of
\begin{equation}\label{buchrad}
r_\Sigma=\frac{9}{4}M,
\end{equation}
which implies upon replacement into Eq.\eqref{junctionex1.1} that $\alpha=9$. This is a major difference between general relativity and more complicated theories like e.g. $f\left(R\right)$, $f\left(R,T\right)$ and hybrid metric-Palatini gravity: in general relativity, one can choose to perform the matching at any radii $r_\Sigma$ as long as $r_\Sigma>2M$ to prevent the collapse to a black-hole, whereas in these theories the extra junction condition $\left[K\right]=0$ forces the matching to be performed at a specific value of $r_\Sigma$. This particular value obtained for $r_\Sigma$ in Eq.\eqref{buchrad} is widely known in the literature and it corresponds to the Buchdahl radius, i.e., the compactness limit for a perfect fluid isotropic star with a non-increasing density. Although this result is not mandatory in GR, it does arise for the particular case of thin-shells satisfying the equation of state of radiation, i.e., $\sigma=2p$. Since we have proven in Sec.\ref{sec:ex1} that in $f\left(R,T\right)$ gravity all spherically symmetric thin-shells will obey the same equation of state, see Eqs.\eqref{sigmapt} and \eqref{sigmaptsca}, our results seem to indicate that the behavior or radiation thin-shells in this theory is consistent with the expected result from GR.

\subsubsection{Matching in the geometrical representation}

First of all, it is essential to verify if the metrics given in Eqs.\eqref{mink1} and \eqref{schw1} are solutions of the field equations in Eq.\eqref{field2} and for which forms of the function $f\left(R,T\right)$. Inserting these metrics into the field equations, one verifies that the function $f\left(R,T\right)$ is constrained to vanish at $R=0$ and $T=0$, i.e., $f\left(0,0\right)=0$. 
 
Let us now analyze the remaining junction conditions in the geometrical representation of the theory. Since both spacetimes present identically vanishing stress-energy tensor $T_{ab}$ and Ricci tensor $R_{ab}$, and consequently vanishing traces $T$ and $R$, the junction conditions $\left[R\right]=0$ and $\left[T\right]=0$ are automatically satisfied. Furthermore, taking the first-order partial derivatives of $R$ and $T$, one verifies that the junction condition $f_{RR}\left[\partial_cR\right]+f_{RT}\left[\partial_cT\right]=0$ is also automatically satisfied independently of the form of the function $f\left(R,T\right)$, as long as $f_{RR}$ and $f_{RT}$ are non-singular at $R=0$ and $T=0$. 

We are thus left with the last junction condition of the set of Eq.\eqref{fullset1}. Since we are dealing with spherically symmetric spacetimes, the analysis of a thin-shell under these conditions is already done in general in Sec.\ref{sec:ex1} and the results for the surface energy density $\sigma$ and transverse pressure $p_t$ of the thin-shell are already given in Eq.\eqref{sigmapt}. Using the spacetimes in Eqs.\eqref{mink1} and \eqref{schw1}, and the matching radius $r_\Sigma$ from Eq.\eqref{buchrad}, we obtain
\begin{equation}\label{sabex1}
\sigma=2p_t=\frac{f_R}{8\pi+f_T}\frac{M}{r_\Sigma^2}\left(\sqrt{1-\frac{2M}{r_\Sigma}}\right)^{-1}=\frac{f_R}{8\pi+f_T}\frac{16}{27M},
\end{equation}
where we have considered $\epsilon=1$ as the normal vector $n^a$ to the hypersurface $\Sigma$ is a spacetime vector pointing in the radial direction. From Eq.\eqref{sabex1}, one concludes that the energy conditions in Eqs.\eqref{nec} to \eqref{dec} will be satisfied whenever $f_R>0$ and $f_T>-8\pi$ or $f_R<0$ and $f_T<-8\pi$. The second of these combinations is not ideal as it inverts the sign of the linear contribution of $R$ in $f\left(R,T\right)$, and thus it shall be discarded. Combining these constraints with $f\left(0,0\right)=0$ previously obtained, one concludes that the function $f\left(R,T\right)$ can be written in a very general form as
\begin{equation}
f\left(R,T\right)=a_1 R+a_2T+\mathcal O\left(2\right),
\end{equation}
where $a_i$ are constants constrained by the inequalities $a_1>0$ and $a_2>-8\pi$, and $\mathcal O\left(2\right)$ collectively denotes all the possible combinations of products of order $2$ or higher in $R$ and $T$, i.e., $a_3 R^2$, $a_4 T^2$, $a_5 RT$, etc., with arbitrary constants $a_i$.

\subsubsection{Matching in the scalar-tensor representation}

To reproduce the previous results in the scalar-tensor representation one must firstly set an explicit form of the function $f\left(R,T\right)$ and compute the corresponding potential. To facilitate the analysis, let us consider one of the simplest forms of the function $f\left(R,T\right)$ for which the scalar-tensor representation is well-defined and the constraint $f\left(0,0\right)=0$ is satisfied, as
\begin{equation}\label{fex1}
f\left(R,T\right)=R+T+\frac{RT}{R_0},
\end{equation}
where $R_0$ is a constant with dimensions of $R$. The second order partial derivatives of Eq.\eqref{fex1} are thus $f_{RR}=f_{TT}=0$ and $f_{RT}=1/R_0$, from which we obtain that the determinant of the matrix $\mathcal M$ defined in Eq.\eqref{matrixeq} is $f_{RR}f_{TT}-f_{RT}^2=R_0^{-2}\neq 0$ and the scalar-tensor representation of the theory is well defined. Thus, we can compute the scalar fields from Eq.\eqref{defsca} which are
\begin{equation}\label{scaex1}
\varphi=1+\frac{T}{R_0}, \qquad \psi=1+\frac{R}{R_0}.
\end{equation}
The relations in Eq.\eqref{scaex1} are invertible, and one can write $R$ and $T$ as functions of $\varphi$ and $\psi$ as $R=R_0\left(\psi-1\right)$ and $T=R_0\left(\varphi-1\right)$. Using Eq.\eqref{defpot}, we can now compute the potential $V\left(\varphi,\psi\right)$ which takes the form
\begin{equation}\label{potex1}
V\left(\varphi,\psi\right)=\frac{RT}{R_0}=R_0\left(\varphi-1\right)\left(\psi-1\right).
\end{equation}
Taking the partial derivatives of Eq.\eqref{potex1} one verifies that $V_\varphi=R_0\left(\psi-1\right)=R$ and $V_\psi=R_0\left(\varphi-1\right)=T$, and so the equations of motion for the scalar fields, i.e., Eqs.\eqref{eomphi} and \eqref{eompsi}, are not only automatically satisfied but also allow us to compute the solutions for the scalar fields in this particular case as $\varphi=1$ and $\psi=1$, both constants. Inserting these considerations into the field equations in Eq.\eqref{fieldst}, one verifies that both metrics in Eqs.\eqref{mink1} and \eqref{schw1} are solutions of these equations. 

Let us now analyze the remaining junction conditions in the scalar-tensor representation. As we have already computed the solutions for the scalar fields $\varphi$ and $\psi$ in the particular choice of $f\left(R,T\right)$ considered and concluded that they are constant, the junction conditions $\left[\varphi\right]=0$ and $\left[\psi\right]=0$ are automatically satisfied. Furthermore, taking the first-order partial derivatives, one also verifies that $\left[\partial_c\varphi\right]=0$ is satisfied.

Finally, we have to settle the last junction condition of Eq.\eqref{fullset3}. Again, the analysis of the stress-energy tensor of the thin-shell is already done in Sec.\ref{sec:ex1} and the resultant surface energy density $\sigma$ and transverse pressure $p_t$ are given in Eq.\eqref{sigmaptsca}. Using the metrics in Eqs.\eqref{mink1} and \eqref{schw1} and $r_\Sigma$ from Eq.\eqref{buchrad} we find
\begin{equation}\label{sab2ex1}
\sigma=2p_t=\frac{\varphi_\Sigma}{8\pi+\psi_\Sigma}\frac{M}{r_\Sigma^2}\left(\sqrt{1-\frac{2M}{r_\Sigma}}\right)^{-1}=\frac{1}{8\pi+1}\frac{16}{27M},
\end{equation}
where we have used $\epsilon=1$ since the normal vector $n^a$ to the hypersurface $\Sigma$ points in the radial direction. The result in Eq.\eqref{sab2ex1} could be obtained directly from Eq.\eqref{sabex1} via the insertion of the particular form of $f\left(R,T\right)$ chosen in Eq.\eqref{fex1}, which confirms the equivalence between the two representations of the theory.

\subsection{A thin-shell surrounding an interior Schwarzschild black-hole}\label{sec:ex3}

Finally, let us consider the matching between two Schwarzschild spacetimes with different masses with a thin-shell at a given hypersurface $\Sigma$ with radius $r_\Sigma$, i.e., this spacetime represents a Schwarzschild black-hole surrounded by a thin-shell, as firstly introduced in Ref.\cite{Brady:1991np}. The interior and exterior spacetimes are described respectively by the line elements
\begin{equation}\label{schwi}
ds^2=-\left(1-\frac{2M_i}{r}\right)dt^2+\left(1-\frac{2M_i}{r}\right)^{-1}dr^2+r^2d\Omega^2,
\end{equation}
\begin{equation}\label{schwe}
ds^2=-\left(1-\frac{2M_e}{r}\right)\alpha dt^2+\left(1-\frac{2M_e}{r}\right)^{-1}dr^2+r^2d\Omega^2,
\end{equation}
where we have considered the usual spherical coordinates $\left(t, r, \theta, \phi\right)$, $\alpha$ is a dimensionless constant introduced to maintain the continuity of the time coordinates, $M_i$ and $M_e$ are the masses of the interior and exterior Schwarzschild spacetimes respectively, and $d\Omega$ is the solid angle line element. As both the spacetimes in Eqs.\eqref{schwi} and \eqref{schwe} are vacuum spacetimes in GR, they feature vanishing stress-energy tensors $T_{ab}=0$, Ricci tensors $R_{ab}=0$, and consequently Ricci scalars $R=0$.

Similarly to the example in Sec.\ref{sec:ex2}, the first junction condition of the sets in Eqs.\eqref{fullset1} and \eqref{fullset3}, i.e., $\left[h_{\alpha\beta}\right]=0$ sets the value of the constant $\alpha$ that guaranteed the continuity of the time coordinate across the hypersurface $\Sigma$, which is in this case
\begin{equation}\label{junctionex3.1}
\alpha=\frac{r_\Sigma-2M_i}{r_\Sigma-2M_e}.
\end{equation} 

As the two regions $\mathcal V^\pm$ are again spherically symmetric, the angular components of the extrinsic curvature coincide, i.e., $K_{\theta\theta}=K_{\phi\phi}$, and the corresponding trace simplifies to $K=K_t^t+2K_\theta^\theta$. The second junction condition in Eqs. \eqref{fullset1} and \eqref{fullset3}, i.e., $\left[K\right]=0$, thus yields
\begin{equation}\label{junctionex3.2}
\left[K\right]=\frac{1}{r^2}\left(\frac{3M_i-2r}{\sqrt{1-\frac{2M_i}{r}}}-\frac{3M_e-2r}{\sqrt{1-\frac{2M_e}{r}}}\right)=0.
\end{equation}
This condition serves as a constraint to set the value of $r$ at which the matching can be performed. In this case however, Eq.\eqref{junctionex3.2} features two roots for $r$, one of which standing in the range $0<r<2M_i$ independently of the value of $M_e$. These solutions are nonphysical, as the thin-shell would stand inside the event-horizon of the interior black-hole and thus one expects the system to undergo full gravitational collapse instead of being in an equilibrium static configuration. We thus discard this solution and keep solely the solution for which the matching occurs ar $r>2M_i$ independently of the value of $M_e$, which is
\begin{equation}\label{matradius}
r_\Sigma=\frac{9}{8}\left[M_e+M_i+\sqrt{\left(M_e-M_i\right)^2+\frac{4}{9}M_e M_i}\right].
\end{equation}
From this solution one verifies that if $M_e>M_i$, the matching must occur at $R>3M_i$, whereas if $M_e<M_i$ the matching will occur at $R<3M_i$. In particular, we verify that if $M_i=0$ we recover the limit obtained in Eq.\eqref{buchrad} for the Martinez shell. Finally, after setting the values of $M_e$ and $M_i$, one can replace the value of $r_\Sigma$ from Eq.\eqref{matradius} into Eq.\eqref{junctionex3.1} to compute the value of $\alpha$.

\subsubsection{Matching in the geometrical representation}

Before proceeding, one should verify if the spacetimes in Eqs.\eqref{schwi} and \eqref{schwe} are solutions of the field equations in Eq.\eqref{field2}. As previously demonstrated in Sec.\ref{sec:ex2}, the Schwarzschild solution is a solution of these field equations as long as the function $f\left(R,T\right)$ satisfies the condition $f\left(0,0\right)=0$.

Similarly to the previous example, both the interior and exterior spacetimes considered are vacuum spacetimes in GR, and thus they present identically vanishing stress-energy tensors $T_{ab}$ and Ricci tensors $R_{ab}$. Taking the traces, one verifies that also $T$ and $R$ vanish identically, and thus the junction conditions $\left[R\right]=0$ and $\left[T\right]=0$ are automatically satisfied. Taking the first-order partial derivatives of $R$ and $T$, one verifies that the junction condition $f_{RR}\left[\partial_cR\right]+f_{RT}\left[\partial_cT\right]=0$ is also automatically satisfied provided that $f_{RR}$ and $f_{TT}$ are non-singular at $R=0$ and $T=0$.

Finally, the last junction condition in the set of Eq.\eqref{fullset1} must be considered. For spherically symmetric spacetimes, this analysis was already conducted in general in Sec.\eqref{sec:ex1} and the results for the surface energy density $\sigma$ and transverse pressure $p_t$ are given in Eqs.\eqref{sigmapt}. For the spacetimes in Eqs.\eqref{schwi} and \eqref{schwe}, these results become
\begin{equation}\label{shellex3}
\sigma=2p_t=\frac{f_R}{r_\Sigma^2\left(f_T+8\pi\right)}\left(\frac{M_e}{\sqrt{1-\frac{2M_e}{r_\Sigma}}}-\frac{M_i}{\sqrt{1-\frac{2M_i}{r_\Sigma}}}\right),
\end{equation}
where $r_\Sigma$ is provided in Eq.\eqref{matradius} and we have considered $\epsilon=1$ as the normal vector $n^a$ to the hypersurface $\Sigma$ is a spacelike vector pointing in the radial direction. From Eqs.\eqref{shellex3} and \eqref{matradius}, one verifies that the energy conditions in Eqs.\eqref{nec} to \eqref{dec} will be satisfied in four different situations. If $M_e>M_i$, then one must have $f_R>0$ and $f_T>-8\pi$ or $f_R<0$ and $f_T<-8\pi$. On the other hand, if $M_e<M_i$, one must have $f_R>0$ and $f_T<-8\pi$ or $f_R<0$ and $f_T>-8\pi$. This results presents a crucial difference with respect to GR or $f\left(R\right)$ theories of gravity. In GR, the choice $M_e<M_i$ prevents completely the energy conditions to be fulfilled. In $f\left(R\right)$, although there are particular forms of the function $f\left(R\right)$ that allow for the the energy conditions to be fulfilled, one needs to impose $f'\left(R\right)<0$, thus effectively inverting the sign of the linear $R$ contribution to the function $f\left(R\right)$. In $f\left(R,T\right)$ however, one can chose $M_e<M_i$ and still satisfy both $f_R>0$ and all the energy conditions, by imposing the constraint $f_T<-8\pi$. Similarly to Sec.\eqref{sec:ex2}, these considerations allow the function $f\left(R,T\right)$ to have the general form
\begin{equation}
f\left(R,T\right)=a_1R+a_2T+\mathcal O\left(2\right),
\end{equation}
where $a_i$ are constants constrained by the inequalities $a_1>0$ and $a_2>-8\pi$ if $M_e>M_i$, or $a_1>0$ and $a_2<-8\pi$ is $M_e<M_i$, and $\mathcal =\left(2\right)$ collectively denotes all possible combinations of products of $R$ and $T$ of order $2$ or higher.

\subsubsection{Matching in the scalar-tensor representation}

Let us now verify that the analysis conducted in the scalar-tensor representation of the theory yields the same results. To do so, one must consider a particular form of the function $f\left(R,T\right)$ and compute the corresponding potential and scalar fields. For simplicity, let us chose the simple form of $f\left(R,T\right)$ for which the scalar-tensor representation is well-defined and the condition $f\left(0,0\right)=0$ is satisfied:
\begin{equation}\label{fex3}
f\left(R,T\right)=a_1 R+a_2T+\frac{RT}{R_0},
\end{equation}
where $a_i$ are arbitrary dimensionless constants and $R_0$ is a constant with units of $R$. The second order derivatives of Eq.\eqref{fex3} are $f_{RR}=f_{TT}=0$ and $f_{TR}=1/R_0$, and thus the determinant of $\mathcal M$ in Eq.\eqref{matrixeq} is $R_0^{-2}\neq0$ and the scalar-tensor representation is well-defined. The scalar fields can thus be computed from Eq.\eqref{defsca} and are
\begin{equation}
\varphi=a_1+\frac{T}{R_0}, \qquad \psi=a_2+\frac{R}{R_0}.
\end{equation}
These relations are invertible and one can write $R=R_0\left(\varphi-a_1\right)$ and $T=R_0\left(\psi-a_2\right)$. Thus, one can compute the potential $V\left(\varphi,\psi\right)$ from Eq.\eqref{defpot}, which is
\begin{equation}
V\left(\varphi,\psi\right)=\frac{RT}{R_0}=R_0\left(\varphi-a_1\right)\left(\psi-a_2\right).
\end{equation}
Taking the partial derivatives of $V\left(\varphi,\psi\right)$ one has $V_\varphi=R_0\left(\psi-a_2\right)=R$ and $V_\psi=R_0\left(\varphi-a_1\right)=T$, and thus the equations of motion for the scalar fields are automatically satisfied and they yield the constant solutions $\varphi=a_1$ and $\psi=a_2$. As a consequence, one is then able to verify that the metrics in Eqs.\eqref{schwi} and \eqref{schwe} are solutions of the field equations in Eq.\eqref{fieldstpf}.

Since the two scalar fields arising from this choice of the function $f\left(R,T\right)$ are constant, the third and fourth junction conditions in Eq.\eqref{fullset3}, i.e., $\left[\varphi\right]=0$ and $\left[\psi\right]=0$ are automatically satisfied. Furthermore, taking the first derivative of $\varphi$ one verifies that the junction condition $\left[\partial_c\varphi\right]=0$ is also automatically satisfied for a constant $\varphi$. 

Finally, considering the last junction condition of Eq.\eqref{fullset3}, and since the spacetimes considered are spherically symmetric, the general results for the surface energy sensity $\sigma$ and the transverse pressure $p_t$ of the thin-shell are already given in Eq.\eqref{sigmaptsca}. Using the metrics in Eqs.\eqref{schwi} and \eqref{schwe}, we obtain
\begin{equation}\label{shellex3st}
\sigma=2p_t=\frac{\varphi_\Sigma}{r_\Sigma^2\left(\psi_\Sigma+8\pi\right)}\left(\frac{M_e}{\sqrt{1-\frac{2M_e}{r_\Sigma}}}-\frac{M_i}{\sqrt{1-\frac{2M_i}{r_\Sigma}}}\right),
\end{equation}
where $r_\Sigma$ is given in Eq.\eqref{matradius} and we have used $\epsilon=1$ since $n^a$ points in the radial direction. Comparing this result with Eq.\eqref{shellex3}, one verifies that Eq.\eqref{shellex3st} could be obtained directly from the previous results simply by introducing the transformation $f_R=\varphi$ and $f_T=\psi$. The constraints on the constants $a_i$ for the energy conditions in Eqs.\eqref{nec} to \eqref{dec} to be satisfied depending on the relationship between $M_e$ and $M_i$ are the same as obtained in the geometrical representation of the theory, i.e., $a_2>0$ and $a_2<-8\pi$ for $M_e<M_i$ or $a_1>0$ and $a_2>-8\pi$ for $M_e>M_i$, which emphasizes the equivalence between the two approaches.

\section{Conclusions}\label{sec:concl}

In this work we have used the distribution formalism to derive the junction conditions of the $f\left(R,T\right)$ theory of gravity not only in the well-known geometrical representation but also in a dynamically equivalent scalar-tensor representation obtained by the introduction of two auxiliary fields. As expected, the generalization of $f\left(R\right)$ gravity to $f\left(R,T\right)$ gives rise to new junction conditions in both representations, which implies that solutions matched in $f\left(R\right)$ may not necessarily be solutions in this theory.

In the geometrical representation, we verified that all the junction conditions previously obtained in Ref. \cite{Senovilla:2013vra} for $f\left(R\right)$ gravity are also present in this theory. However, the fact that the field equations depend explicitly in $T$ and its partial derivatives (via the differencial terms in $f_R$) leads to two extra junction conditions: the trace of the stress-energy tensor of the thin-shell must vanish, i.e., $S=0$, and the trace of the stress-energy tensor $T_{ab}$ must be continuous at the separation, i.e., $\left[T\right]=0$. The first of these conditions forces the terms proportional to $\left[\partial_cR\right]$ and $\left[\partial_c\right]$ to cancel in $S_{ab}$ and thus, unlike $f\left(R\right)$, the thin-shell is completely described by the discontinuity of the extrinsic curvature $K_{ab}$. For the particular case of smooth matching, one recovers that the extrinsic curvature and the partial derivatives of $R$ must be continuous similarly to $f\left(R\right)$, which also forces the partial derivatives of $T$ to be continuous.

If one considers the scalar-tensor representation instead, one verifies that the action that describes the theory is similar to a Brans-Dicke action with two scalar fields with a parameter $\omega_{\text{BD}}=0$ and a potential depending on both fields. Since this scalar-tensor representation is only defined when the Hessian matrix of the function $f\left(R,T\right)$ is invertible, which corresponds also to an invertibility of the functions $\varphi\left(R,T\right)$ and $\psi\left(R,T\right)$, then one verifies that the scalar fields $\varphi$ and $\psi$ must be continuous. This result emphasizes the difference between the scalar-tensor representations of $f\left(R\right)$ and respective extensions, of which $f\left(R,T\right)$ is an example, and Brans-Dicke theories with a potential, as in the latter the continuity of the scalar fields only arises if the scalar fields feature a kinetic term, i.e., when $\omega_{\text{BD}}\neq 0$. The junction condition for the trace $S=0$ of the thin-shell forces the partial derivatives of the scalar field $\varphi$ to be continuous and one recovers the dependency of $S_{ab}$ solely in the lump of the extrinsic curvature $\left[K_{ab}\right]$. For smooth matching, one recovers the continuity of the extrinsic curvature and the partial derivatives of the scalar field $\psi$, thus proving the equivalence between the two representations.

If the function $f\left(R,T\right)$ is at most second order in $R$, and first order in $RT$, some of the junction conditions previously obtained van be discarded, namely the continuity of $R$ and $T$. This happens because the terms from which these junction conditions arise are proportional to second-order derivatives of $f\left(R,T\right)$ which vanish in this particular case. Unlike in $f\left(R\right)$, it was shown that these particular cases do not lead to the appearance of gravitational double-layers at the separation hypersurface due to the condition $S=0$, which imposes a relationship between $\left[\partial_cR\right]$ and $\left[\partial_cT\right]$, as well as $\left[R\right]$ and $\left[T\right]$, which effectively cancels the terms associated to the double-layers. The same behavior is found in the scalar-tensor representation of the theory, in which one is able to discard the junction condition $\left[\psi\right]=0$ for the corresponding potential $V\left(\varphi,\psi\right)$ without giving rise to gravitational double-layers.

Nevertheless, there are still particular forms of the function $f\left(R,T\right)$ for which gravitational double-layers arise, associated to forms of the function at most quadratic in $R$ and linear in $T$ without crossed products. These cases again allow for $R$ and $T$ to be discontinuous and all the quantities related to the double-layer, i.e., energy and momentum fluxes, tangential stresses, and the double-layer distribution function, depend explicitly on $\left[R\right]$ and $\left[T\right]$. These particular cases do not have any counterparts in the scalar-tensor representation of the theory as, for the forms of the function $f\left(R,T\right)$ necessary to conduct this analysis, the scalar-tensor representation is not well-defined ad the determinant of the Hessian matrix of $f\left(R,T\right)$ vanishes identically.

It is particularly interesting that the extra junction condition forcing the trace of the stress-energy tensor of the thin-shell to vanish has important consequences in spacetimes with spherically symmetric thin-shells. In particular, these thin-shells are described by an equation of state of radiation, i.e., $\sigma=2p_t$. Consequently, all the energy conditions (null, weak, strong and dominant) will be satisfied simultaneously if the surface energy density of the shell is positive, and violated simultaneously otherwise. This effectively eases the analysis of physically relevant thin-shells in this framework, as one can focus solely in guaranteeing that $\sigma$ is positive, without the need to fine-tune the parameters in an attempt to verify the energy conditions for $\sigma$ and $p_t$ simultaneously, as it happens in other theories e.g. hybrid metric-Palatini gravity \cite{rosaworm}.

Another possible application of the junction conditions derived in this work is the construction of thin-shell wormhole solutions \cite{visser1, visser2} via the truncation of two black-hole solutions at a given radius $r>r_h$, where $r_h$ is the radius of the event horizon. The junction condition $\left[K\right]=0$ might prevent this procedure to be applicable to Schwarzschild black-holes, but it is expectable that this task can be fulfilled with charged black-holes \cite{Eiroa:2015hrt}. Furthermore, it has been shown that the condition $\left[K\right]=0$ is discarded in the Palatini approach to $f\left(R\right)$ gravity \cite{Olmo:2020fri}, which simplifies the construction of thin-shell wormholes \cite{Lobo:2020vqh}. In this sense, it would be of major importance to compute the junction conditions of the $f\left(R,T\right)$ gravity in the Palatini formalism.

\begin{acknowledgments}
We would like to thank Francisco S. N. Lobo for the comments and suggestions. This work was supported by the European Regional Development Fund and the programme Mobilitas Pluss (MOBJD647).
\end{acknowledgments}



\begin{thebibliography}{99}

\bibitem{darmois}
G. Darmois, ``Les equations de la gravitation einsteinienne'',
Memorial des sciences mathematiques, {\bf 25}, 565 (1927).

\bibitem{lichnerowicz}
A. Lichnerowicz,
{\it Th\'eories Relativistes de la Gravitation et de l'Electromagn\'etisme}
(Masson, Paris 1955).

\bibitem{Israel:1966rt} 
W.~Israel, ``Singular hypersurfaces and thin shells in general
relativity'', Nuovo Cim.\ B {\bf 44}, 1 (1966).

\bibitem{papa} A. Papapetrou and A. Hamoui, ``Couches simples de
mati\'ere en relativit\'e g\'en\'erale'', Annales de l'Institute Henri
Poincar\'e A, A. l'I. H. P. A, {\bf 9}, 179 (1968).

\bibitem{taub} A. H. Taub,
``Space-times with distribution valued curvature tensors'',
Journal of Mathematical Physics {\bf 21}, 1423 (1980).

\bibitem{oppenheimer}
J. R. Oppenheimer and H. Snyder, ``On continued gravitational 
contraction'', Phys. Rev. \textbf{56}, 455 (1939).

\bibitem{senovilla1}
F. Fayos, J. M. M. Senovilla, and R. Torres, ``General matching of 
two spherically symmetric spacetimes'', Phys. Rev. D \textbf{54},
4862 (1996).

\bibitem{lanczos1}
K. Lanczos, ``Bemerkungen zur de Sitterschen Welt'',
Physikalische Zeitschrift \textbf{23}, 539-547, (1922).

\bibitem{lanczos2}
K. Lanczos, ``Fl\"achenhafte verteiliung der Materie in der Einsteinschen
Gravitationstheorie'', Annalen der Physik (Leipzig) \textbf{74},
518-540, (1924).

\bibitem{Martinez:1996ni} 
E.~A.~Martinez,
``Fundamental thermodynamical equation of a self-gravitating system'',
Phys.\ Rev.\ D {\bf 53}, 7062 (1996);
arXiv:gr-qc/9601037.

\bibitem{Lemos:2017mci} 
  J.~P.~S.~Lemos, M.~Minamitsuji, and O.~B.~Zaslavskii,
``Thermodynamics of extremal rotating thin shells in an
extremal BTZ spacetime and the extremal black hole entropy'',
  Phys.\ Rev.\ D {\bf 95}, no. 4, 044003 (2017);
arXiv:1701.02348 [hep-th].

\bibitem{Lemos:2017aol} 
J.~P.~S.~Lemos, M.~Minamitsuji and O.~B.~Zaslavskii,
``Unified approach to the entropy of an extremal
rotating BTZ black hole: Thin shells and horizon limits'',
Phys.\ Rev.\ D {\bf 96}, no. 8, 084068 (2017);
[arXiv:1709.08637 [hep-th]].

\bibitem{Lemos:2015ama} 
J.~P.~S.~Lemos, G.~M.~Quinta and O.~B.~Zaslavskii,
``Entropy of an extremal electrically charged thin shell and the extremal black hole'',
  Phys.\ Lett.\ B {\bf 750}, 306 (2015);
  arXiv:1505.05875 [hep-th].

\bibitem{Lemos:2016pyc} 
  J.~P.~S.~Lemos, G.~M.~Quinta and O.~B.~Zaslavskii,
  ``Entropy of extremal black holes: Horizon limits through charged
   thin shells in a unified approach'',
  Phys.\ Rev.\ D {\bf 93}, no. 8, 084008 (2016)
  [arXiv:1603.01628 [hep-th]].
  
\bibitem{brito}  
R. Brito, V. Cardoso, J. V. Rocha, "Interacting shells in AdS
spacetime and chaos", Phys. Rev. D \textbf{94}, 024003, (2016).

\bibitem{rosafluid}
J. L. Rosa, P. Pi\c{c}arra, "Existence and stability of relativistic fluid
spheres supported by thin shells", Phys. Rev. D \textbf{102}, 064009
(2020).

\bibitem{Vignolo:2018eco} 
  S.~Vignolo, R.~Cianci and S.~Carloni,
  ``On the junction conditions in $f(R)$-gravity with torsion'',
  Class.\ Quant.\ Grav.\  {\bf 35}, no. 9, 095014 (2018)
  [arXiv:1801.08344 [gr-qc]].  

\bibitem{Senovilla:2013vra} 
  J.~M.~M.~Senovilla,
  ``Junction conditions for F(R)-gravity and their consequences'',
  Phys.\ Rev.\ D {\bf 88}, 064015 (2013)
  [arXiv:1303.1408 [gr-qc]].
  
\bibitem{Deruelle:2007pt} 
  N.~Deruelle, M.~Sasaki and Y.~Sendouda,
  ``Junction conditions in f(R) theories of gravity'',
  Prog.\ Theor.\ Phys.\  {\bf 119}, 237 (2008)
  [arXiv:0711.1150 [gr-qc]].
  
\bibitem{Olmo:2020fri}
G.~J.~Olmo and D.~Rubiera-Garcia, ``Junction conditions in Palatini $f(R)$ gravity,''
Class. Quant. Grav. \textbf{37} (2020) no.21, 215002
doi:10.1088/1361-6382/abb924
[arXiv:2007.04065 [gr-qc]].
  
\bibitem{Barrabes:1997kk} 
  C.~Barrabes and G.~F.~Bressange,
  ``Singular hypersurfaces in scalar - tensor theories of gravity'',
  Class.\ Quant.\ Grav.\  {\bf 14}, 805 (1997)
  [gr-qc/9701026].
  
\bibitem{suffern}
K.G. Suffern, "Singular hypersurfaces in the Brans-Dicke theory of
gravity", Journal of Physics A: Mathematical and General
\textbf{15}, 1599 (1982).

\bibitem{Davis:2002gn} 
  S.~C.~Davis,
  ``Generalized Israel junction conditions for a Gauss-Bonnet brane world'',
  Phys.\ Rev.\ D {\bf 67}, 024030 (2003)
  [hep-th/0208205].

\bibitem{rosaworm}

J. L. Rosa, J. P. S. Lemos, F. S. N. Lobo, "Wormholes in generalized hybrid metric-Palatini gravity obeying the matter null energy condition everywhere", Phys. Rev. D \textbf{98}, 064054 (2018).

\bibitem{cardoso}

V. Cardoso, P. Pani, "Testing the nature of dark compact objects: a status report", Living Rev. Relativ. \textbf{22} 4 (2019).

\bibitem{clifton}

T. Clifton, P. G. Ferreira, A. Padilla, C. Skordis, “Modified gravity and cosmology”, Phys. Rep. \textbf{513}, 1
(2012).

\bibitem{nojiri}

S. Nojiri, S. D. Odintsov, “Unified cosmic history in modified gravity: from f(R) theory to Lorentz
non-invariant models”, Phys. Rep. \textbf{505}, 59 (2011).

\bibitem{Nojiri:2017ncd}
  S.~Nojiri, S.~D.~Odintsov and V.~K.~Oikonomou,
  ``Modified Gravity Theories on a Nutshell: Inflation, Bounce and
Late-time Evolution,''
  Phys.\ Rept.\  {\bf 692} (2017) 1
  doi:10.1016/j.physrep.2017.06.001
  [arXiv:1705.11098 [gr-qc]].

\bibitem{perlmutter}

S. Perlmutter \textit{et al.} (Supernova Cosmology Project Collaboration), "Measurements of Omega and Lambda from 42 high redshift supernovae", Astrophys. J. \textbf{517}, 565 (1999).

\bibitem{riess}

A. G. Riess \textit{et al.} (Supernova Search Team Collaboration), "Observational evidence from supernovae for an accelerated universe and a cosmological constant", Astron. J. \textbf{116}, 1009 (1998).

\bibitem{sotiriou}

T. P. Sotiriou and V. Faraoni, “f(R) theories of gravity”, Rev. Mod. Phys. \textbf{82}, 451 (2010).

\bibitem{felice}

A. De Felice and S. Tsujikawa, “f(R) theories”, Living Rev. Rel. \textbf{13}, 3 (2010)

\bibitem{bohmer1}

C. G. B\"ohmer, T. Harko, and F. S. N. Lobo, “Dark matter as a geometric effect in f(R) gravity”, Astropart. Phys. \textbf{29}, 386 (2008).

\bibitem{bohmer2}

C. G. B\"ohmer, T. Harko, and F. S. N. Lobo, “Generalized virial theorem in f(R) gravity”, J. Cosmol. Astropart. Phys. (JCAP) \textbf{03} (2008) 024.

\bibitem{khoury1}

J. Khoury and A. Weltman, “Chameleon fields: Awaiting surprises for tests of gravity in space”, Phys. Rev. Lett. \textbf{93}, 171104 (2004).

\bibitem{khoury2}

J. Khoury and A. Weltman, “Chameleon cosmology”, Phys. Rev. D \textbf{69}, 044026 (2004).

\bibitem{harko}

T. Harko, F. S. N. Lobo, S. Nojiri, S. D. Odintsov, "$f\left(R,T\right)$ gravity", Phys. Rev. D \textbf{84}, 024020 (2011).

\bibitem{zaregonbadi}

R. Zaregonbadi, M. Farhoudi, N. Riazi, "Dark matter from $f\left(R,T\right)$ gravity", Phys. Rev. D \textbf{94}, 084052 (2016).

\bibitem{velten}

H. Velten, T. R. P. Caram\^es, "Cosmological inviability of $f\left(R,T\right)$ gravity", Phys. Rev. D \textbf{95}, 123536 (2017).

\bibitem{houndjo1}

M. J. S. Houndjo, “ Reconstruction of f(R, T) gravity describing matter dominated and accelerated
phases”, Int. J. Mod. Phys. D. \textbf{21}, 1250003 (2012).

\bibitem{houndjo2}

M. J. S. Houndjo and O. F. Piattella, “Reconstructing f(R, T) gravity from holographic dark energy”,
Int. J. Mod. Phys. D. \textbf{21}, 1250024 (2012).

\bibitem{jamil}

M. Jamil, D. Momeni, M. Reza and R. Myrzakulov, "Reconstruction of some cosmological models in $f\left(R,T\right)$ cosmology", Euro. Phys. J. C \textbf{72}, 1999 (2012).

\bibitem{alvarenga}

F. G. Alvarenga, M. J. S. Houndjo, A. V. Monwanou, J. B. C. Orou, "Testing some f(R,T) gravity models from energy conditions", Journal of Modern Physics \textbf{4}, 130-139 (2013).

\bibitem{wu}

J. Wu, G. Li, T. Harko, S. D. Liang, "Palatini formulation of $f\left(R,T\right)$ gravity theory, and its cosmological implications", Euro. Phys. J. C \textbf{78}, 430 (2018).

\bibitem{carvalho}

G. A. Carvalho, R. V. Lobato, P. H. R. S. Moraes, J. D. V. Arba\~{n}il, E. Otoniel, R. M. Marinho Jr, M. Malheiro, "Stellar equilibrium configurations of white dwarfs in the $f\left(R,T\right)$ gravity", The European Physical Journal C volume \textbf{77}, 871 (2017).

\bibitem{deb}

D. Deb, F. Rahaman, S. Ray, B.K. Guha, "Strange stars in $f\left(R,T\right)$ gravity", JCAP03 044 (2018).

\bibitem{maurya}

S.K. Maurya, A. Errehymy, D. Deb, F. Tello-Ortiz, M. Daoud, "Study of anisotropic strange stars in $f\left(R,T\right)$ gravity: An embedding approach under the simplest linear functional of the matter-geometry coupling", Phys. Rev. D \textbf{100}, 044014 (2019).

\bibitem{bhatti}

M. Z. Bhatti, Z. Yousaf, M. Yousaf, "Stability of self-gravitating anisotropic fluids in $f\left(R,T\right)$ gravity", Physics of the Dark Universe \textbf{28} 100501 (2020).

\bibitem{ordines}

T. M. Ordines, E. D. Carlson, "Limits on $f\left(R,T\right)$ gravity from Earth’s atmosphere", Phys. Rev. D \textbf{99}, 104052 (2019).

\bibitem{sahoo}

P. Sahoo, P.H.R.S. Moraes, M. M. Lapola, P.K. Sahoo, "Traversable wormholes in the traceless $f\left(R,T\right)$ gravity", 	arXiv:2012.00258 [gr-qc].

\bibitem{mishra}

A. K. Mishra, U. K. Sharma, V. C. Dubey, A. Pradhan, "Traversable wormholes in $f\left(R,T\right)$ gravity", Astrophysics and Space Science \textbf{365}, 34 (2020).

\bibitem{moraes}

P. H. R. S. Moraes, P. K. Sahoo, "Modeling wormholes in $f\left(R,T\right)$ gravity", Phys. Rev. D \textbf{96}, 044038 (2017).

\bibitem{banerjee}

A. Banerjee, M. K. Jasim, S. G. Ghosh, "Traversable wormholes in $f\left(R,T\right)$ gravity satisfying the null energy condition with isotropic pressure", arXiv:2003.01545 [gr-qc].

\bibitem{rosacosmo}

J. L. Rosa, S. Carloni, J. P. S. Lemos, F. S. N. Lobo, "Cosmological solutions in generalized hybrid metric-Palatini gravity", Phys.Rev.D \textbf{95} 12, 124035 (2017).

\bibitem{rosabrane}

J. L. Rosa, D. A. Ferreira, D. Bazeia, F. S. N. Lobo, "Thick brane structures in generalized hybrid metric-Palatini gravity", Eur. Phys. J. C \textbf{81} 1, 20 (2021).

\bibitem{rosasudden}

J. L. Rosa, F. S. N. Lobo, D. Rubiera-Garcia, "Sudden singularities in generalized hybrid metric-Palatini cosmologies", [arXiv:2103.02580 [gr-qc]] (2021).

\bibitem{bohmer3}

N. Tamanini, C. G. Bohmer, "Generalized hybrid metric-Palatini gravity", Physical Review D\textbf{87}, 084031 (2013).

\bibitem{bombacigno}

F. Bombacigno, F. Moretti, G. Montani, "Scalar modes in extended hybrid metric-Palatini gravity: weak field phenomenology", Phys. Rev. D \textbf{100}, 124036 (2019).

\bibitem{Brady:1991np}
P.~R.~Brady, J.~Louko and E.~Poisson, ``Stability of a shell around a black hole'', Phys. Rev. D \textbf{44}, 1891 (1991).

\bibitem{visser1}

M. Visser, "Traversable wormholes: Some simple examples", Phys. Rev. D \textbf{39}, 3182 (1989).

\bibitem{visser2}

M. Visser, "Traversable wormholes from surgically modified Schwarzschild spacetimes", Nucl. Phys. B \textbf{328}, 11 203 (1989).

\bibitem{Eiroa:2015hrt}
E.~F.~Eiroa and G.~Figueroa Aguirre, "Thin-shell wormholes with charge in F(R) gravity",
Eur. Phys. J. C \textbf{76} no.3, 132 (2016).

\bibitem{Lobo:2020vqh}
F.~S.~N.~Lobo, G.~J.~Olmo, E.~Orazi, D.~Rubiera-Garcia and A.~Rustam, ``Structure and stability of traversable thin-shell wormholes in Palatini $f(\mathcal{R})$ gravity,''
Phys. Rev. D \textbf{102}, 104012 (2020).


\end{thebibliography}
\end{document}